\documentclass[aps,prb,twocolumn,floats,fleqn,epsfig]{revtex4}
\usepackage{epsfig,psfrag,subfigure,amsopn}
\usepackage{graphicx,psfrag,subfigure}
\usepackage{amsmath,amssymb,amsbsy}

\usepackage{color}
\usepackage{ulem}
\usepackage{verbatim} 
\usepackage{setspace}
\usepackage{wrapfig}
\usepackage{bbold}
\usepackage{float}

\newcommand\beq{\begin{equation}}
\newcommand\eeq{\end{equation}}
\newcommand\bea{\begin{eqnarray}}
\newcommand\eea{\end{eqnarray}}
\newcommand\bsq{\begin{subequations}}
\newcommand\esq{\end{subequations}}

\newcommand\non{\nonumber}
\newcommand\noi{\noindent}

\newcommand\ig{\includegraphics}
\newcommand\bib{\bibitem}

\newcommand\ga{\gamma}
\newcommand\de{\delta}
\newcommand\De{\Delta}

\newcommand\si{\sigma}

\newcommand\lam{\lambda}
\newcommand\pa{\partial}
\newcommand\la{\langle}
\newcommand\ra{\rangle}

\newcommand\dg{\dagger}
\newcommand\ua{\uparrow}
\newcommand\da{\downarrow}

\newcommand\vk{{\vec k}}

\begin{document}

\title{Electron dynamics in graphene with spin-orbit couplings and periodic 
potentials}

\author{Ranjani Seshadri and Diptiman Sen}
\affiliation{\small{Centre for High Energy Physics, Indian Institute of 
Science, Bengaluru 560 012, India}} 

\date{\today}

\begin{abstract}
We use both continuum and lattice models to study the energy-momentum 
dispersion and the dynamics of a wave packet for an electron moving in 
graphene in the presence of spin-orbit couplings and either a single 
potential barrier or a periodic array of potential barriers. Both Kane-Mele 
and Rashba spin-orbit couplings are considered. A number of special things 
occur when the Kane-Mele and Rashba couplings are equal in magnitude.
In the absence of a potential, the dispersion then consists of both 
massless Dirac and massive Dirac states. A periodic potential is known to 
generate additional Dirac points; we show that spin-orbit couplings generally 
open gaps at all those points, but if the two spin-orbit couplings are equal, 
some of the Dirac points remain gapless. We show that the massless and
massive states respond differently to a potential barrier; the massless 
states transmit perfectly through the barrier at normal incidence while the 
massive states reflect from it. In the presence of a single 
potential barrier, we show that there are states localized along the 
barrier. Finally, we study the time evolution of a wave packet in the 
presence of a periodic potential. We discover special points in momentum 
space where there is almost no spreading of a wave packet; there are six 
such points in graphene when the spin-orbit couplings are absent.
\end{abstract}

\maketitle

\section{Introduction}

Graphene has been the arena for an enormous amount of experimental and 
theoretical research for several years~\cite{been08,neto09,rev3,rev4,rev5}. 
Graphene consists of a two-dimensional hexagonal lattice of $sp^2$ hybridized 
carbon atoms in which the $\pi$ electrons hop between nearest neighbors. The 
energy spectrum is gapless at two points in the 
Brillouin zone; these points are labeled as $K$ and $K'$ (this is called 
the valley degree of freedom), and the energy-momentum dispersion around 
those points has the Dirac form $E_{\vk} = \hbar v |\vk|$, where $v \simeq 
10^6 ~m/s$ is the Fermi velocity. The Dirac nature of the electrons is
responsible for many interesting properties of graphene, such as the quantum 
Hall effect~\cite{novoselov05,zhang05}, Klein tunneling through a 
barrier~\cite{katsnelson06}, effects of crossed electric and magnetic 
fields~\cite{lukose07}, unusual transport properties of superconducting 
graphene junctions~\cite{beenakker1,bhattacharjee06,bhattacharjee07,beenakker2,
maiti07}, multichannel Kondo physics \cite{ksen3,kond1,kond2,kond3,kond4}, 
interesting power laws in the local density of states near an 
impurity~\cite{cheianov06,mariani07,bena08,bena09}, and atomic collapse in 
the presence of charged impurities~\cite{levitov07,wang13}. The effects of 
Kane-Mele and Rashba spin-orbit (SO) 
interactions~\cite{kane1,kane2,rashba,zarea09,bena14} on the 
impurity-induced local density of states and on transport across barriers 
have been examined~\cite{seshadri1}, and the effect of Rashba SO couplings on 
tunneling through $pn$ and $pnp$ junctions has been studied~\cite{liu12}.
SO couplings may be induced in graphene in various ways, such as 
a transverse electric field~\cite{gmitra09}, adatom deposition~\cite{weeks11},
or proximity to a three-dimensional topological insulator such as $\rm Bi_2 
Se_3$~\cite{kou13,zhang14}, or functionalizing with methyl~\cite{zollner15}. 
(We note that the Kane-Mele and Rashba SO couplings are respectively referred 
to as intrinsic and extrinsic SO couplings in the literature; however, 
in this paper we will refer to them as Kane-Mele and Rashba couplings 
for convenience). The dynamics of wave packets in graphene has been 
studied in a number of papers using both the microscopic lattice model of 
graphene~\cite{costa} and a continuum theory which is valid close to 
the Dirac points~\cite{maksi,singh14,singh15}.

Recently it has been analytically shown that applying a potential in
graphene which is periodic in one or both coordinates can produce additional 
Dirac points~\cite{park,brey,barbier2,barbier3};
%analytically~\cite{park,brey,barbier1,barbier2,barbier3,masir1,masir2} 
experimental evidence for this in transport measurements has been presented 
in Ref.~\onlinecite{dubey} although an alternative explanation has been 
proposed in Ref.~\onlinecite{drienovsky}. On 
the other hand, a potential which is independent of one coordinate and is a 
random function of the other coordinate is known to give rise to 
supercollimation, namely, a wave packet moves only in the direction in which 
the potential varies randomly~\cite{choi}. 

In this paper, we study the energy dispersion and wave packet dynamics 
in graphene in the presence of a periodic potential and SO couplings. The 
plan of the paper is as follows. In Sec. II, we use a continuum theory 
near the Dirac points (labeled $K$ and $K'$) to study the energy dispersion 
in the presence of SO couplings and a periodic array of $\de$-function
potentials. In the absence of a periodic potential, a 
Kane-Mele SO coupling produces a gap at the Dirac points which is doubly 
degenerate (for a given momentum) due to the spin and valley degrees of 
freedom. A combination of Kane-Mele and Rashba SO couplings produces four 
non-degenerate states. When the two SO couplings are equal, two of the states 
have a gapless Dirac form while the other two have a gapped Dirac form. The 
presence of a periodic potential generates additional Dirac points as known 
in the literature; we show that spin-orbit couplings generally open gaps at 
those points unless the two couplings are equal. (A related study was 
carried out in Ref.~\onlinecite{shakouri}).
In Sec. III, we use the microscopic lattice model of graphene to study 
the energy dispersion in the presence of SO couplings and a single potential 
barrier. This confirms the results obtained using continuum theory in 
Sec. II. In addition, we show that there are states which are localized 
along the barrier and whose energies lie in the bulk gap~\cite{seshadri2}.
In Sec. IV, we use the lattice model to study wave packet dynamics in 
the presence of a periodic potential and SO couplings. The wave packets are 
be taken to be Gaussians. For graphene without any SO couplings, 
we analytically find six special points in the Brillouin zone where there is 
negligible spreading of a wave packet. When the Kane-Mele and Rashba SO 
couplings are non-zero but equal, we show that wave packets constructed 
from the two kinds of states (the gapless Dirac and the gapped Dirac states
discussed in Sec. II) respond quite differently to the barriers. We 
conclude in Sec. V with a summary of our main results.

\section{Continuum theory around Dirac points}

In this section, we will use a continuum theory around the Dirac points 
(which lie at two momenta called $K$ and $K'$) to study the
energy spectrum in the presence of SO couplings and a potential which is 
periodic in one direction. We will consider both a Kane-Mele SO 
coupling~\cite{kane1,kane2} called $\De_{KM}$ and a Rashba SO 
coupling~\cite{rashba,zarea09,bena14}
called $\lam_R$. Further, a periodic potential which only depends on the 
$y$-coordinate is applied; the precise form of this potential will be 
specified below, and we will assume that it has the symmetry $V(y) = V(-y)$. 
Since the system has translational symmetry along the $x$ direction, the 
momentum $k_x$ along this direction is a good quantum number. 
The complete Hamiltonian close to the Dirac points is then given by
\bea H &=& \hbar v_F (\tau^z \si^x k_x + i \si^y \frac{\pa}{\pa y}) ~+~ 
\De_{KM} \tau^z \si^z s^z \non \\
&& + ~\lam_R ( \tau^z\si^x s^y - \si^y s^x) ~+~ V(y), 
\label{ham1} \eea
where $\si^a$, $\tau^a$ and $s^a$ are Pauli matrices corresponding to 
sublattice ($\si^z = +(-) ~1$ for $A(B)$), valley ($\tau^z = +(-) ~1$ for 
$K(K')$) and spin ($s^z = +(-) ~1$ for up (down) spin) respectively. The Fermi
velocity $v_F \simeq 10^6 ~m/s$ and $k_x$ is the deviation from the Dirac 
point. (Henceforth we will set $\hbar = 1$ unless otherwise mentioned).

We first look at the various symmetries of the Hamiltonian in Eq.~\eqref{ham1};
these will imply certain symmetries of the energy spectrum and eigenstates. 

\noi 1. For a given value of $\tau^z$, we have 
\beq H(k_x,y,\tau^z) ~=~ \si^x s^y ~H(k_x,-y,\tau^z) ~\si^x s^y. \label{sym1} 
\eeq 

\noi 2. The Hamiltonians at $K$ and $K'$ are related by
\beq H(k_x,y,\tau^z) ~=~ \tau^x \si^x s^z ~H(k_x,-y,-\tau^z) ~\tau^x \si^x s^z.
\label{sym2} \eeq

\noi 3. $\tau^x \si^y$, $\tau^y \si^y$ and $\tau^z$ all commute with $H$ and 
anticommute with one another. As a result, the $\tau^z = \pm 1$ sectors are degenerate.

\noi 4. If $\lam_R = 0$, the Hamiltonian has the symmetry
\beq H(k_x,y,\tau^z) = \si^x s^x ~H(k_x,-y,\tau^z) ~\si^x s^x. \label{sym3} 
\eeq

\noi 5. For a given value of $\tau^z$, the Hamiltonian has the symmetry
\beq H(k_x,y,\tau^z) = \si^y s^x ~H(-k_x,y,\tau^z) ~\si^y s^x. \label{sym4}
\eeq
This implies that the energy spectrum is invariant under $k_x \to - k_x$.

We observe that the symmetries in Eqs.~\eqref{sym1} and \eqref{sym2} flip 
the spin $s^z \to - s^z$; we therefore get a double degeneracy of all 
energy levels due to spin.

At the Dirac point $K$, i.e. $\tau^z = +1$, Eq.~\eqref{ham1} reduces to 
a $4 \times 4$ matrix given by
\bea H &=& v_F (\si^x k_x + i \si^y \frac{\pa}{\pa y}) ~+~ \De_{KM} \si^z s^z 
\non \\
&& +~ \lam_R (\si^x s^y - \si^y s^x) ~+~ V(y). \eea
For a periodic potential satisfying $V(y) = V(y+d)$, the eigenstates can be 
labeled by a Bloch momentum $\chi_y$ (which lies in the range $[-\pi/d,
\pi/d]$), namely, $\psi (k_x,y+d) = e^{i \chi_y d} \psi (k_x,y)$.
The symmetry $\si^x s^y H(k_x,-y) \si^x s^y = H(k_x,y)$ then implies that 
the spectrum is symmetric about $\chi_y=0$ for all $k_x$.

\begin{figure}[H]
\begin{center}
%\subfigure{\includegraphics[width=7.1cm]{periodic.eps}}
\subfigure{\includegraphics[width=7.1cm]{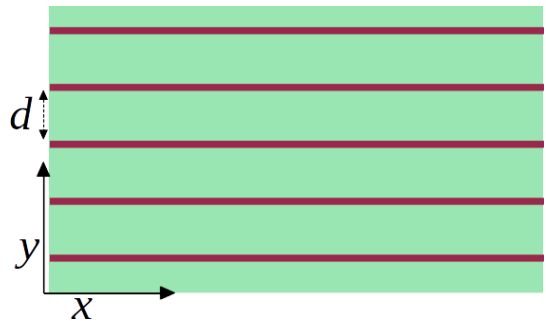}}
\end{center}
\caption{Schematic diagram of a $\de$-function potential which repeats 
periodically in the $y$ direction with a spacing $d$.} \label{schematic} 
\end{figure}

We will numerically compute the energy spectrum for a periodic $\de$-function
potential which is independent of the $x$ coordinate, i.e.,
\beq V(x,y) ~=~ C ~\sum_{n=-\infty}^{\infty} \de (y-nd), \label{defn1} \eeq
where $C$ is the strength of the $\de$-function. ($C$ has the dimensions of 
energy times length). The unit cell size of the periodic potential in the 
$y$ direction is $d$. A schematic picture of the periodic potential is 
shown in Fig.~\ref{schematic}. 

One way of studying the effect of a 
$\de$-function potential in a Dirac Hamiltonian is to note that it induces a 
discontinuity in the wave function of the form $\psi(x,y=d+) = \exp 
[i (C/\hbar v_F) \si^y] \psi(x,y=d-)$ for a $\de$-function of strength $C$ 
located at $y=d$.~\cite{deb12} Using this along with the Bloch theorem which 
states that $\psi (x,y=d+) = \exp (i \chi_y d) \psi (x,y=0+)$, where $\chi_y$
is the Bloch momentum, we can reduce the problem of finding the energies and
eigenstates as a function of $k_x$ and $\chi_y$ to solving a differential 
equation within a single unit cell of the periodic potential. Namely, if we 
write $\psi (x,y) = e^{i k_x x} f(y)$, then the four-component spinor $f(y)$ 
must satisfy 
\bea && [v_F (\si^x k_x + i \si^y \frac{\pa}{\pa y}) ~+~ \De_{KM} \si^z s^z 
\non \\ 
&& +~ \lam_R (\si^x s^y - \si^y s^x)] ~f ~=~ E ~f \eea
in the region $0 < y < d$, subject to the boundary condition $e^{i \chi_y d} 
f(0+) = e^{i (C/\hbar v_F) \si^y} f(d-)$. However we found that this method
is numerically not convenient for finding the energy dispersion.

We have therefore used a different numerical method for finding the dispersion.
Given a value of $k_x$ and $\chi_y$, the general wave function consistent
with the Bloch theorem is given by
\beq \psi (x,y) = e^{i k_x x} ~\sum_{m=-\infty}^\infty ~e^{i (\chi_y + 2 \pi m
/d) y} f_m (y). \label{psixy} \eeq
Let us truncate the range of $m$ in Eq.~\eqref{psixy} to go from $-q$ to $+q$;
this gives a total of $2q+1$ bands. The Hamiltonian in this basis is then a 
$4 (2q+1)$- dimensional 
matrix with blocks of matrix elements as follows. First, there are $2q+1$ 
blocks on the diagonal which are given by $4 \times 4$ matrices of the form
\beq v_F [\si^x k_x - \si^y (\chi_y + \frac{2 \pi m}{d})] + \De_{KM} \si^z s^z
+ \lam_R (\si^x s^y - \si^y s^x). \eeq
Second, the identity 
\beq C \sum_{n=-\infty}^\infty ~\de (y-nd) ~=~ \frac{C}{d} ~
\sum_{m=-\infty}^\infty ~e^{i 2 \pi m y/d}, \eeq
implies that between any two blocks labeled by $m$ and $m'$ (each label runs
from $-q$ to $+q$, and $m, ~m'$ may or may not be equal), there will be a 
coupling given by $(C/d) I_4$, where $I_4$ is the $4 \times 4$ identity 
matrix. Putting these together we get the total Hamiltonian from which we
can obtain $4 (2q+1)$ energy levels.

\begin{figure}[h!]
\begin{center}
\subfigure[]{\includegraphics[width=4.2cm]{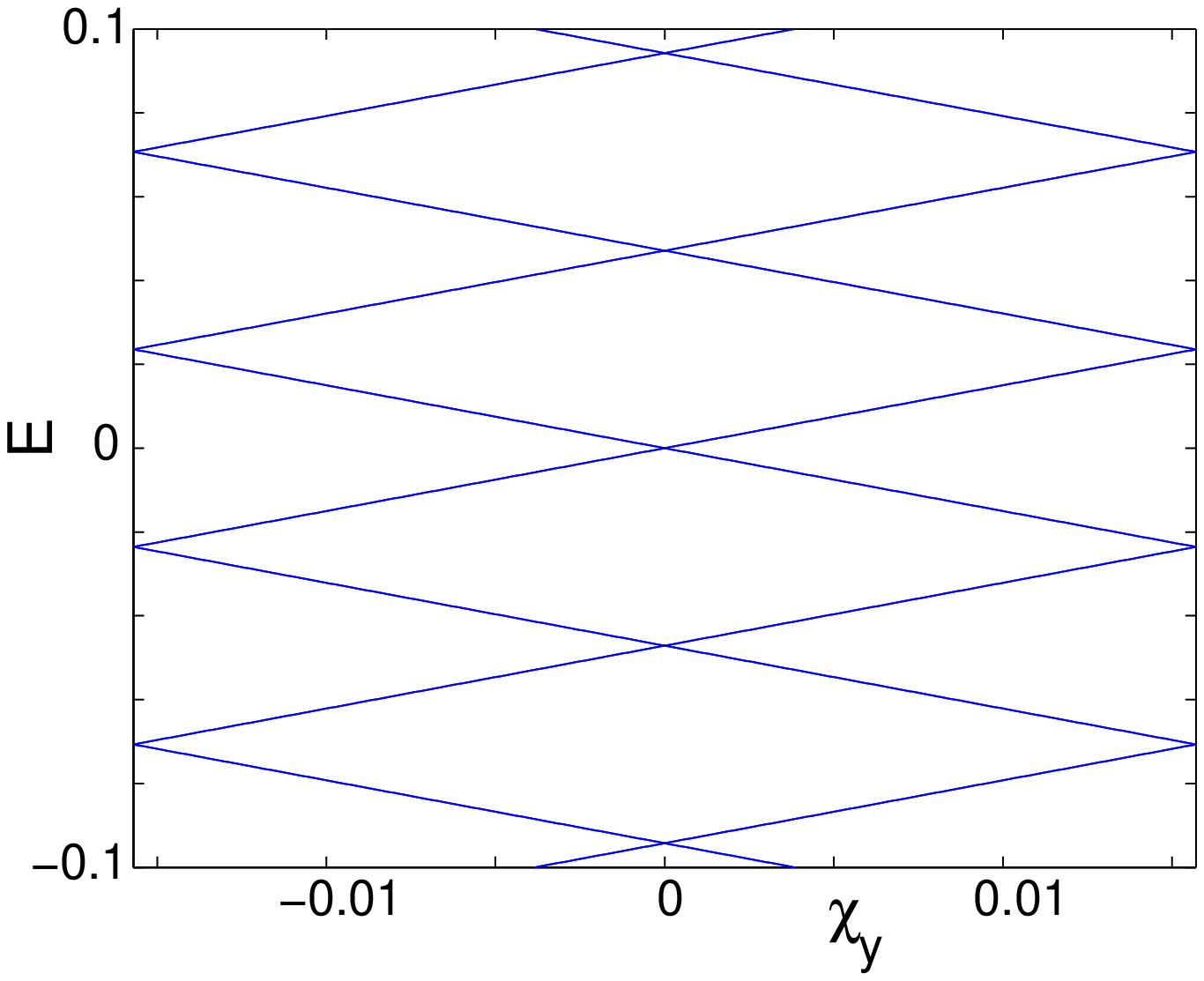}}
\subfigure[]{\includegraphics[width=4.2cm]{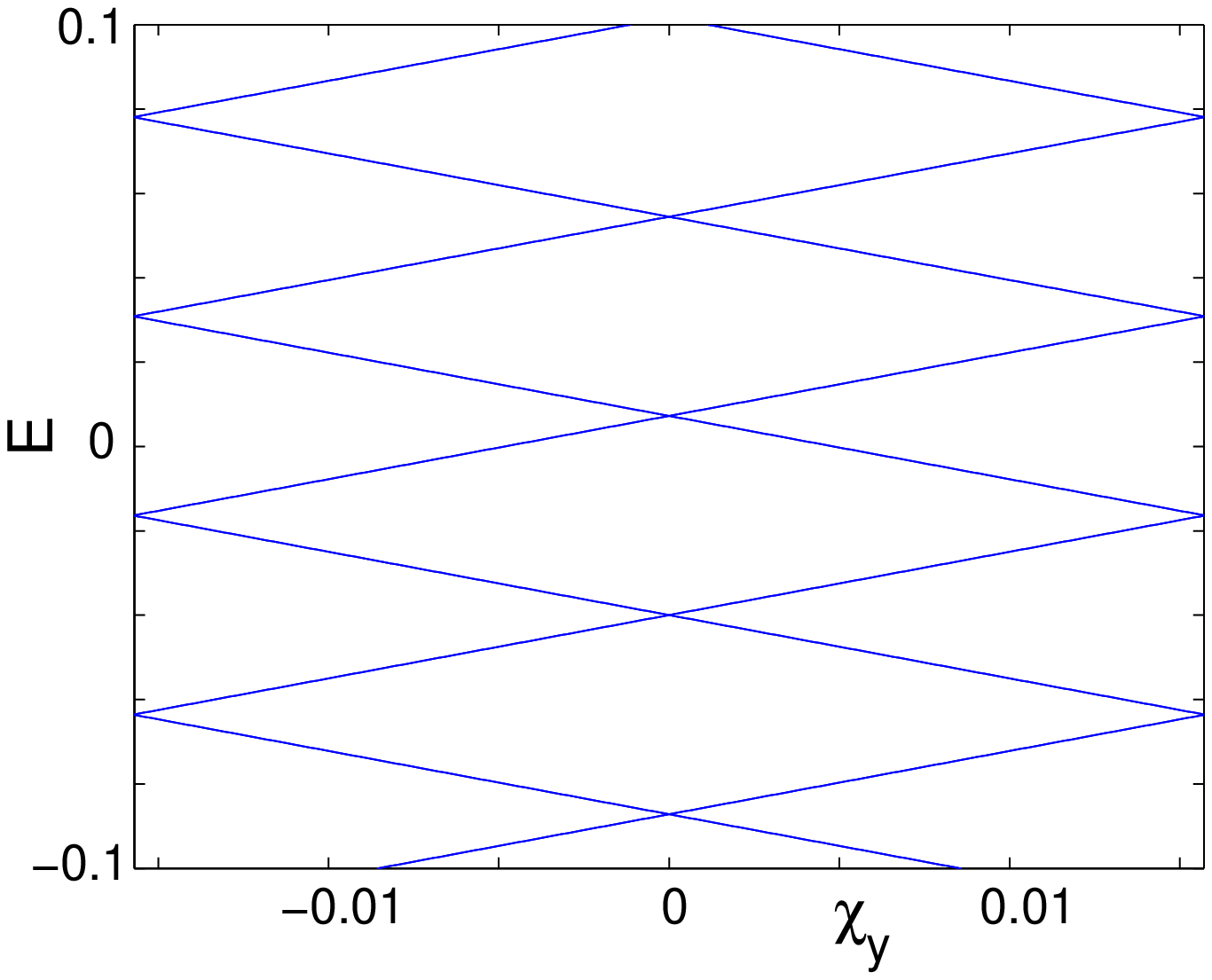}} \\
%\vskip -0.2 cm
\subfigure[]{\includegraphics[width=4.2cm]{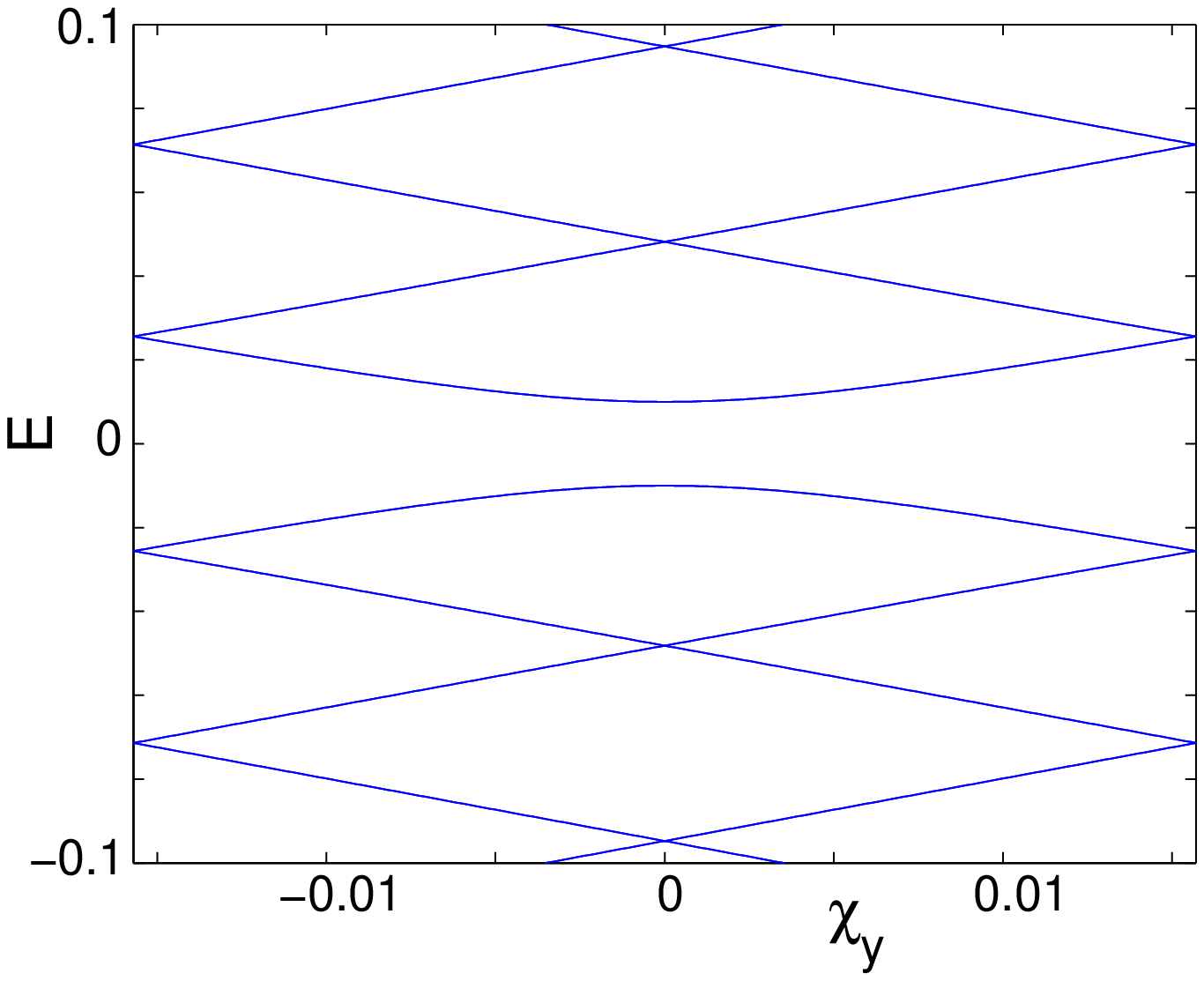}}
\subfigure[]{\includegraphics[width=4.2cm]{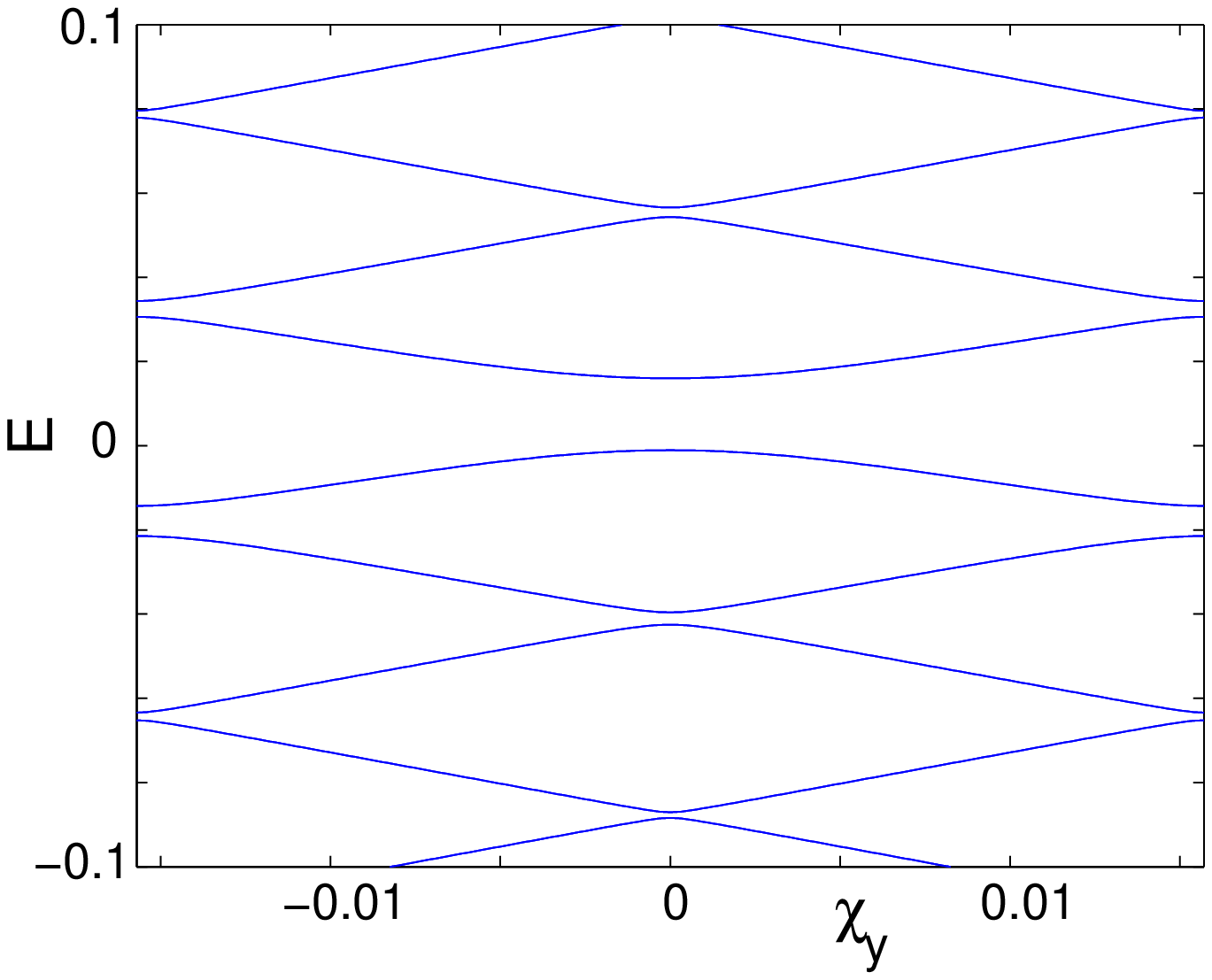}} \\
%\vskip -0.2 cm
\subfigure[]{\includegraphics[width=4.2cm]{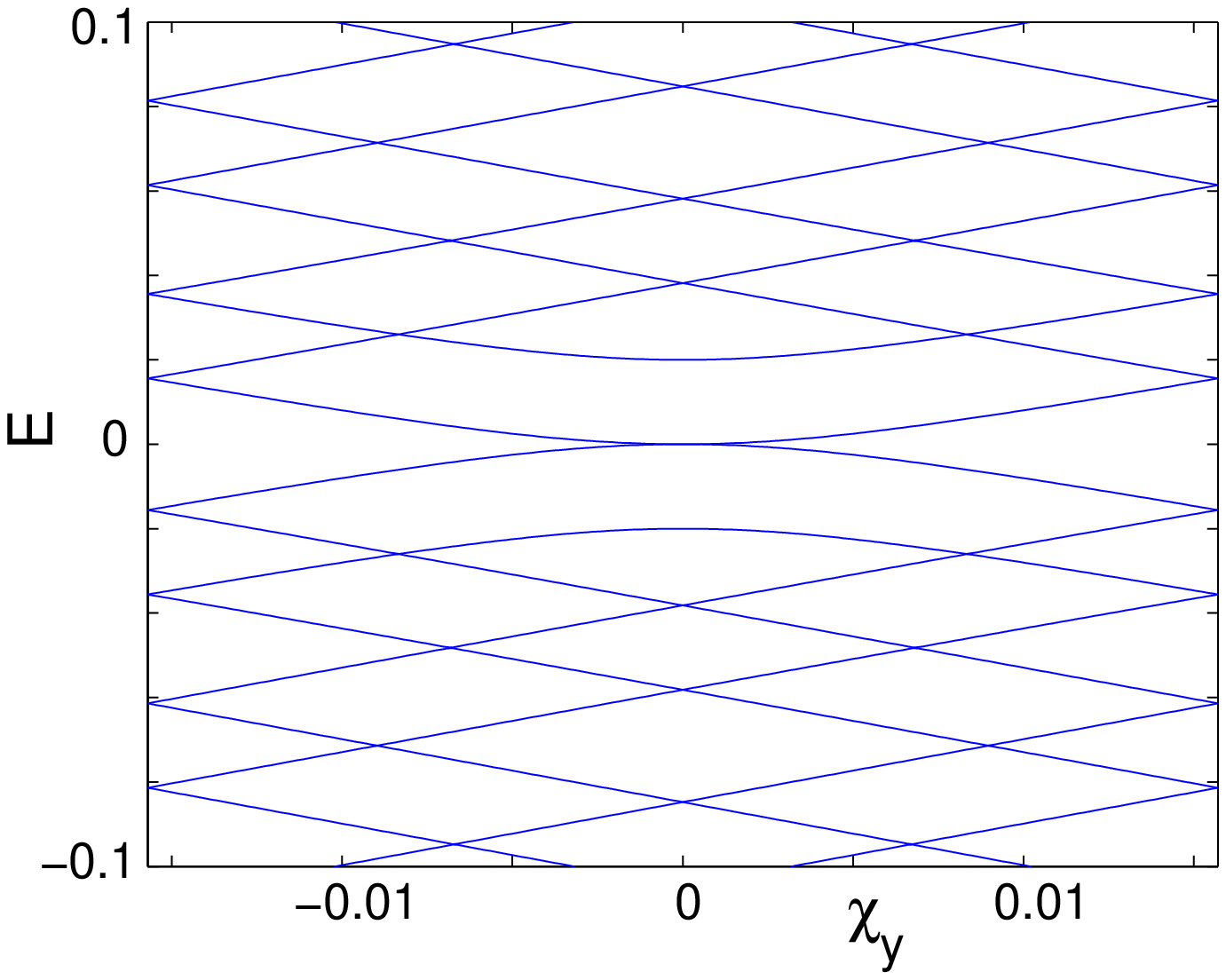}}
\subfigure[]{\includegraphics[width=4.2cm]{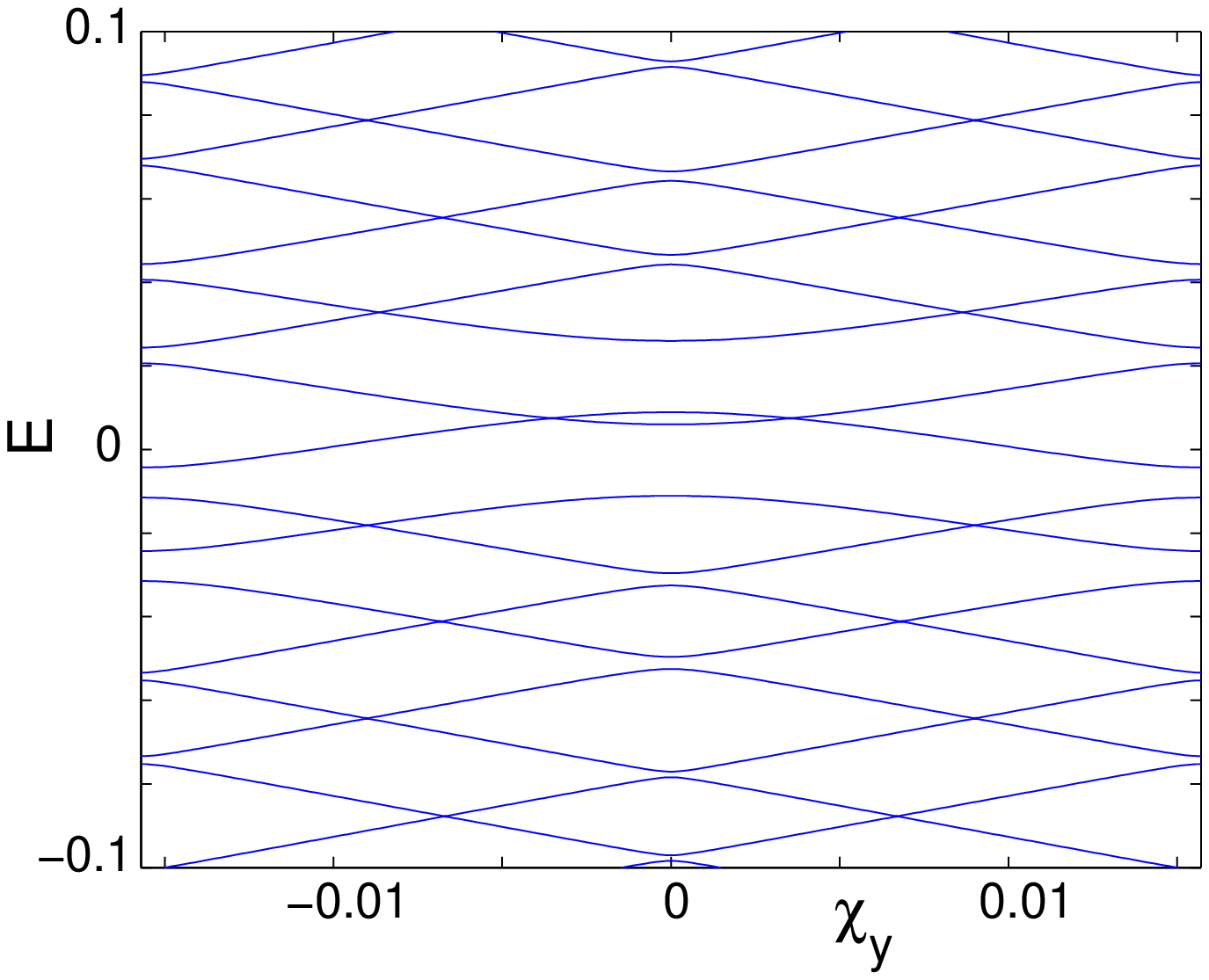}} \\
%\vskip -0.2 cm
\subfigure[]{\includegraphics[width=4.2cm]{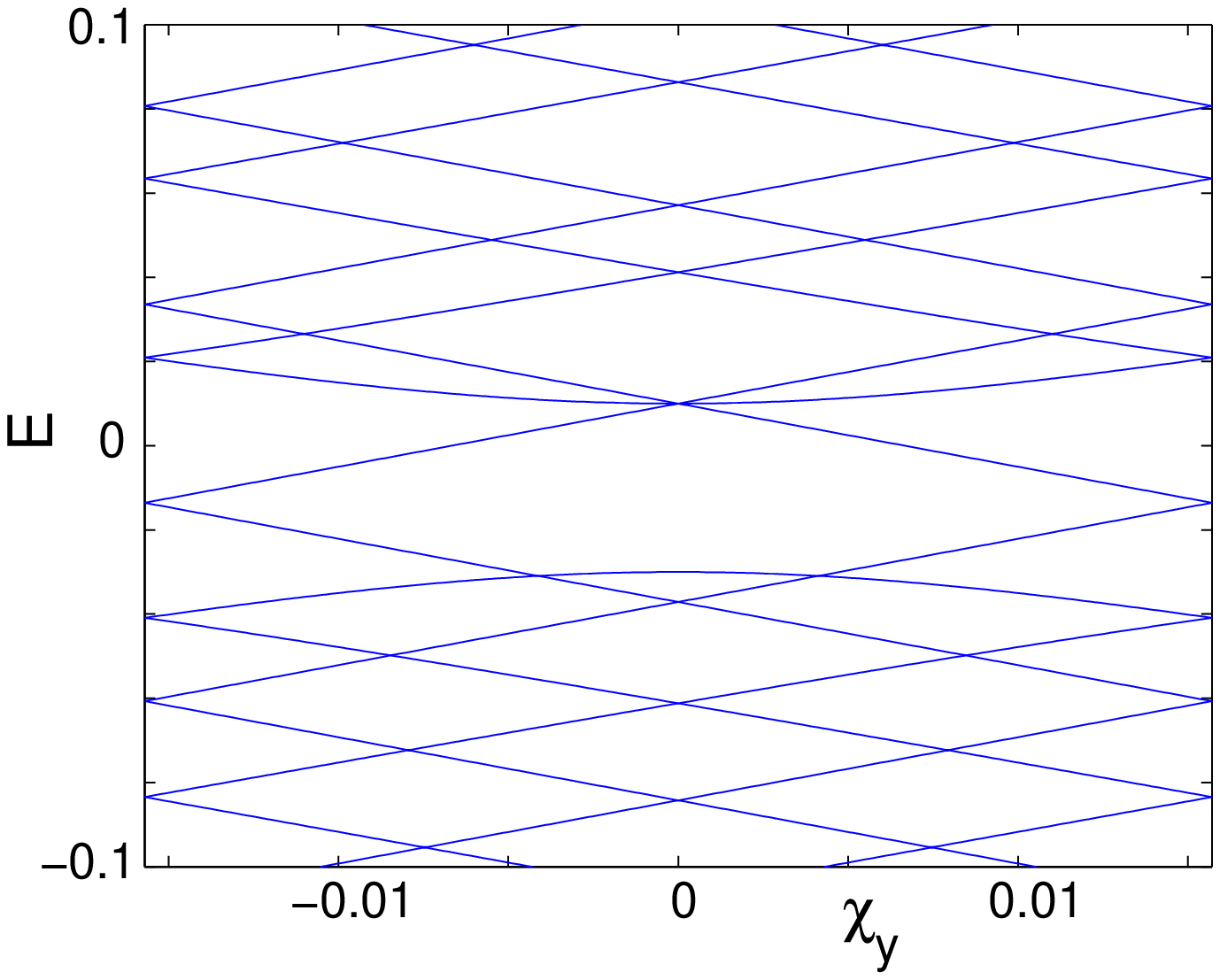}}
\subfigure[]{\includegraphics[width=4.2cm]{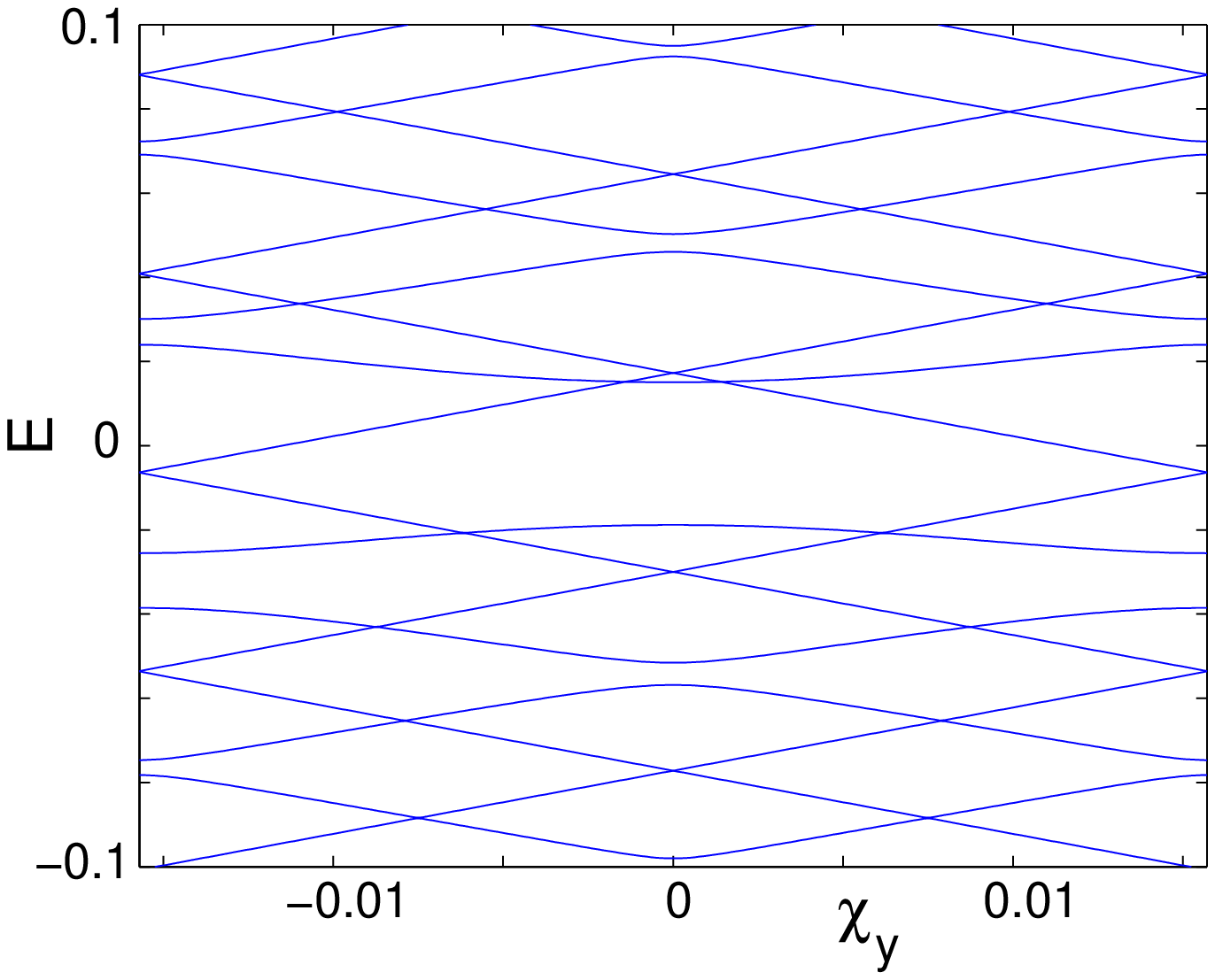}} \\
\caption{$E$ vs $\chi_y$ for $k_x = 0$ and $d=200$. ($d$ is in units of the 
lattice spacing, while $E$ is in units of $\ga$). The figures on the left 
((a),(c),(e),(g)) are in the absence of the $\de$-function periodic 
potential ($C=0$), while the ones on the right ((b),(d),(f),(h)) are with 
$C=\pi \ga /2$ times the lattice spacing. In (a) and (b), $\De_{KM} = \lam_R 
= 0$; in (c) and (d), $\De_{KM} = 0.01 ~\ga$, $\lam_R 
= 0$; in (e) and (f), $\De_{KM} =0$, $\lam_R = 0.01 ~\ga$; in (g) and (h), 
$\De_{KM} = \lam_R = 0.01 ~\ga$.} \label{fig01} \end{center} \end{figure}

In Fig.~\ref{fig01}, we show the energy spectrum $E$ versus the Bloch momentum 
$\chi_y$ (lying in the range $[-\pi/d, \pi/d]$) for $\tau^z =1$, $k_x =0$ and 
$d=200$, and various values of the $\de$-function strength $C$ and SO 
couplings $\De_{KM}$ and $\lam_R$. (In these calculations, we have kept 41 
bands, namely, $q=20$. We have checked that the results do not change 
noticeably if we consider more than 41 bands). To see 
the effects of the periodic potential clearly, we have shown the spectra 
without the potential in Figs.~\ref{fig01} (a), (c), (e) and (g),
and with the potential in Figs. ~\ref{fig01} (b), (d), (f) and (h).
Fig.~\ref{fig01} (b) shows that for graphene without any SO couplings 
($\De_{KM} = \lam_R = 0$), additional gapless Dirac points appear at the center
($\chi_y=0$) and the ends of the reduced Brillouin zone ($\chi_y = \pm \pi/d$)
when a periodic potential is present. We can understand the appearance
of these gapless Dirac points as follows. For normal incidence on
a barrier (i.e., for $k_x = 0$), a gapless Dirac particle transmits
perfectly (this is called Klein tunneling). The absence of reflection
implies that the periodic potential does not lead to any mixing between 
modes with momenta $k_y = +m \pi/d$ and $- m \pi/d$. (Recall that a potential
with periodicity $d$ can only produce scattering between pairs of states whose
$y$-momenta differ by an integer multiple of $2 \pi /d$). Hence the energy 
degeneracy between the modes at $k_y = \pm m \pi/d$ remains unbroken, and no 
gap is produced. Next, Figs.~\ref{fig01} (d) and (f) show the effects of 
Kane-Mele and Rashba SO couplings separately; we see that these couplings
generally
open gaps at the additional Dirac points. Finally, Fig.~\ref{fig01} (h) shows 
that when both SO couplings are present with $\De_{KM} = \lam_R$, some of the 
gapless Dirac points are restored; these gapless points are particularly easy 
to see at the ends of Brillouin zone ($\chi_y = \pm \pi/d$). We will now see 
why $\De_{KM} = \pm \lam_R$ is special.

When the potential $V(y)=0$, the momenta $k_x$ and $k_y$ are both good 
quantum numbers. The energy spectrum of the Hamiltonian in Eq.~\eqref{ham1} 
is then given by~\cite{seshadri1}
\beq \Big[E^2 ~-~ v_F^2 \vk^2 ~-~ \De_{KM}^2 \Big]^2 ~=~ 4 \lam_R^2
~ (E - \De_{KM})^2, ~~~~~~~~~~~ \label{enrel1} \eeq
where $\vk^2 = k^2_x + k^2_y$. This can be solved to give four branches of 
solutions for $E$,
\bea E &=& \lam_R ~\pm~ \sqrt{v_F^2 \vk^2 ~+~ (\De_{KM} - 
\lam_R)^2},~~~ \text{and} \non \\
E &=& - \lam_R ~\pm~ \sqrt{v_F^2 \vk^2~+~ (\De_{KM} + \lam_R)^2}.
\label{enrel2} \eea
We therefore see that if $\De_{KM} = \pm \lam_R$, the dispersion in the 
region around $(k_x,k_y)=(0,0)$ has the massless Dirac form in two of the 
branches ($E = \pm v_F |\vec{k}|$ plus a constant) and the massive Dirac 
form in the other two branches ($E = \pm \sqrt{v_F^2 \vk^2 + 4 \De_{KM}^2}$ 
plus a constant). Depending on which branch we consider, we expect two 
different kinds of behaviors when a periodic potential $V(y)$ is turned on:
additional gapless Dirac points and perfect Klein tunneling at normal 
incidence ($k_x = 0$) from the massless Dirac branches, and gaps at the
additional Dirac points and a non-zero reflection from the massive 
Dirac branches. This can be shown as follows.

For $k_x=0$, Eq.~\eqref{ham1} takes the form
\bea H &=& i v_F \si^y \frac{\pa}{\pa y} + \De_{KM} \tau^z \si^z s^z + 
\lam_R (\tau^z \si^x s^y - \si^y s^x) \non \\
&& +~ V(y). \label{ham2} \eea
This Hamiltonian commutes with $\tau^z$ and $\si^y s^x$; we can therefore work
in a particular sector of eigenstates of $\tau^z$ and $\si^y s^x$ with 
eigenvalues equal to $+1$ or $-1$. Since $(\tau^z \si^z s^z) (\tau^z \si^x 
s^y) = \si^y s^x$, we see that the combination $\De_{KM} \tau^z \si^z s^z + 
\lam_R \tau^z \si^x s^y$ vanishes in the sector $\si^y s^x = -1$ if $\De_{KM}
= \lam_R$ and in the sector $\si^y s^x = +1$ if $\De_{KM} = - \lam_R$.
In these sectors, therefore, the Hamiltonian in \eqref{ham2} reduces to 
\beq H ~=~ i v_F \si^y \frac{\pa}{\pa y} ~\pm~ \lam_R ~+~ V(y), \label{ham3} 
\eeq
where the $\pm$ signs in front of $\lam_R$ are for the cases $\De_{KM} = \pm 
\lam_R$ respectively; these are the sectors which contain the massless Dirac 
modes if $k_x = 0$ and $V(y) = 0$.
Next, we find that for an arbitrary potential $V(y)$, the eigenstates
and spectrum of Eq.~\eqref{ham3} are given by
\bea \psi_{k_y,s} (y) &=& \exp [ ik_y y + (is/v_F) \int_0^y dy' V(y')] ~
u_{k_y,s}, \non \\
E_{k_y} &=& \pm \lam_R ~-~ v_F s k_y, \label{tran} \eea
where the spinor 
$u_{k_y,s}$ is an eigenstate of $\si^y$ with eigenvalue $s= \pm 1$
and an eigenstate of $\si^y s^x$ with eigenvalue $\pm 1$. We thus see that
there is perfect transmission through any potential $V(y)$, and the
spectrum varies linearly with $k_y$. For a periodic potential, the perfect 
transmission and hence the absence of reflection for the massless Dirac 
modes implies that the degeneracy between states at $k_y = \pm m \pi/d$ 
remains unbroken, and no gap is produced at the additional Dirac points.
In Sec. IV, we will see directly that the massless and massive 
Dirac states indeed show different transmission and reflection properties.

\section{Lattice model}

\begin{figure}[h]
\begin{center}
%\subfigure{\includegraphics[width=7.6cm]{edited_lattice_zoom.eps}}
\subfigure{\includegraphics[width=7.6cm]{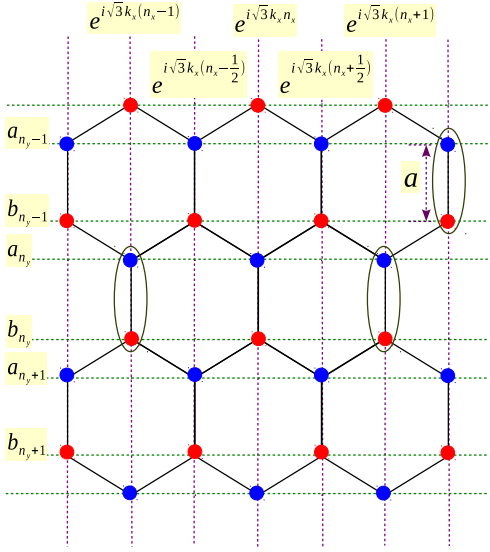}}
\end{center}
\caption{Schematic diagram of the honeycomb lattice used to calculate the
spectra shown in Fig.~\ref{lattice_disp}. The $x$ coordinate increases
from left to right while the $y$ coordinate increases from top to bottom.
The plane wave factors shown at the top are dependent on the momentum $k_x$.
The unit cells are denoted by ellipses and consist of one $a$ site and 
one $b$ site.} \label{hexlat} \end{figure}

In this section we use the microscopic lattice model to study the energy 
spectrum in the presence of a periodic potential and SO couplings. We will 
consider the honeycomb lattice shown in Fig.~\ref{hexlat} with periodic 
boundary conditions in both directions. (We will usually set the 
nearest-neighbor lattice spacing $a = 0.142$ nm equal to 1). The zigzag 
rows run parallel to the $x$ direction. Each unit cell consists of an $a$ 
site and a $b$ site; the cells are labeled by two integers $(n_x,n_y)$ as 
shown. (The size of a unit cell in the $y$ direction is $3a/2$). Since the 
system has translational symmetry along the $x$ direction, the momentum 
$k_x$ in that direction is a good quantum number. The plane wave factors 
depending on $k_x$ are shown at the top of Fig.~\ref{hexlat}.

In second quantized notation, the complete Hamiltonian $H$ of the lattice 
model is the sum of four terms,
\begin{subequations}
\bea H_0 &=& -\ga ~\sum_{\la ij \ra,s} ~f^{\dg}_{is} f_{js}, \label{nosoc} \\
H_{KM} &=& it_2 ~\sum_{\la \la ij \ra \ra,s} ~\nu_{ij} ~s^z_{ss} f^{\dg}_{is}
f_{js}, \label{KM} \\
H_R &=& i\lam_R ~\sum_{\la ij \ra,ss'} ~\hat{z} \cdot (\vec{d}_{ij}
\times \vec{s})_{ss'} ~f^{\dg}_{is} f_{js'}, \label{Rashba} \\
H_V &=& \sum_i ~V_i ~f^{\dg}_i f_i, \label{potential} \\
H &=& H_0 ~+~ H_{KM} ~+~ H_R ~+~ H_V, \label{hamtot} \eea \end{subequations}
where $\la i j \ra$ denotes nearest neighbors ($i$ and $j$ are labeled
by $n_x, n_y$), $\la \la i j \ra \ra$ denotes next-nearest neighbors, and 
the subscripts $s,s'$ denote the spin component $s^z = \pm 1$. 
Eqs.~\eqref{nosoc}, \eqref{KM} and \eqref{Rashba} describe graphene without 
any SO couplings, with Kane-Mele \cite{kane1,kane2} and with Rashba SO terms 
respectively~\cite{rashba,zarea09,bena14}. In \eqref{nosoc}, $\ga \simeq 
2.8$ ~eV
denotes the nearest-neighbor hopping amplitude; the Fermi velocity in Sec. II 
is given by $v_F = (3/2) \ga a$. In \eqref{KM}, $\nu_{ij} = \pm 1$ depending 
on the relative orientation of the two successive nearest-neighbor vectors 
which join site $j$ to its next-nearest-neighbor site $i$.
In \eqref{Rashba}, $\vec{d}_{ij}$ denotes the vector joining 
the nearest-neighbor sites $i$ and $j$. In the continuum theory near the
Dirac points, Eq.~\eqref{KM} reduces to the Kane-Mele term in Eq.~\eqref{ham1}
with $\De_{KM} = 3 \sqrt{3} t_2$, while Eq.~\eqref{Rashba} reduces to the
Rashba term in Eq.~\eqref{ham1}. Finally, we will take the potential $V_i$ in
\eqref{potential} to be a periodic function of $n_y$ and independent of $n_x$.
More specifically, we will choose the periodic potential $V_i$ to be 
composed of Gaussians, rather than the $\de$-function potentials that we 
considered in Sec. II. Namely, we will take
\beq V (n_y) ~=~ \frac{C}{\si \sqrt{2\pi}} ~\sum_{n=-\infty}^\infty ~
e^{-[(3a/2) n_y - nd]^2/(2 \si^2)}, \eeq
where $d$ is the periodicity of the potential; in our calculations we 
have chosen the width of the Gaussians to be $\si = 4a$. 

{} From the Hamiltonian in Eq.~\eqref{hamtot}, we can write down the 
eigenvalue equations for an energy $E$. For a given momentum $k_x$ we can 
effectively reduce the system to a one-dimensional chain which runs along the 
$y$-direction. The unit cells of the chain are labeled by an integer $n_y$; 
each unit cell has four variables labeled $a_{n_y\ua},~a_{n_y\da}, ~b_{n_y\ua}$
and $b_{n_y\da}$. Using the plane wave factors shown in Fig.~\ref{hexlat}, we 
find the following equations. 
\begin{widetext}
\bea
E ~a_{n_y\ua} &=& -~ \ga ~\Big\{ b_{n_y\ua} ~+~ 2 \cos(\frac{\sqrt{3}k_x}{2}) 
b_{n_y-1\ua} \Big\} ~+~ 2t_2 ~\Big\{ \sin( \sqrt{3}k_x)a_{n_y\ua} ~-~\sin( 
\frac{\sqrt{3}k_x}{2}) (a_{n_y-1\ua} ~+~ a_{n_y+1\ua}) \Big\} \non \\
&& +~ i \lam_R ~\Big\{ \Big( \cos(\frac{\sqrt{3}k_x}{2}) ~+~ \sqrt{3} \sin 
(\frac{\sqrt{3}k_x}{2} ) \Big) b_{n_y-1\da}-b_{n_y\da}\Big\} ~+~ V_{n_y} ~
a_{n_y\ua}, \non \\
E ~a_{n_y\da} &=& -~ \ga ~\Big\{ b_{n_y\da} ~+~ 2 \cos(\frac{\sqrt{3}k_x}{2})
b_{n_y-1\da} \Big\} ~-~ 2 t_2 ~\Big\{ \sin( \sqrt{3}k_x)a_{n_y\da} ~-~
\sin(\frac{\sqrt{3}k_x}{2}) (a_{n_y-1\da} ~+~ a_{n_y+1\da}) \Big\} \non \\
&& +~ i \lam_R ~\Big\{ \Big( \cos(\frac{\sqrt{3}k_x}{2}) ~-~ \sqrt{3} \sin 
(\frac{\sqrt{3}k_x}{2} ) \Big)b_{n_y-1\ua} ~-~ b_{n_y\ua}\Big\} ~+~ V_{n_y} ~
a_{n_y\da}, \non \\
E ~b_{n_y\ua} &=& -~ \ga ~\Big\{ a_{n_y\ua} ~+~ 2 \cos(\frac{\sqrt{3}k_x}{2}) 
a_{n_y+1\ua} \Big\} ~- ~2 t_2 ~\Big\{ \sin( \sqrt{3}k_x)b_{n_y\ua} ~-~ \sin
(\frac{\sqrt{3}k_x}{2}) (b_{n_y-1\ua} ~+~ b_{n_y+1\ua}) \Big\} \non \\
&& -~ i \lam_R ~\Big\{ \Big( \cos(\frac{\sqrt{3}k_x}{2}) ~-~ \sqrt{3} \sin 
(\frac{\sqrt{3}k_x}{2} ) \Big) a_{n_y+1\da} ~-~ a_{n_y\da}\Big\} ~+~ V_{n_y} ~
b_{n_y\ua}, \non \\
E ~b_{n_y\da} &=& -~ \ga ~\Big\{ a_{n_y\da} ~+~ 2 \cos(\frac{\sqrt{3}k_x}{2}) 
a_{n_y+1\da} \Big\} ~+ ~2 t_2 ~\Big\{ \sin( \sqrt{3}k_x)b_{n_y\da} ~-~ \sin
(\frac{\sqrt{3}k_x}{2}) (b_{n_y-1\da} ~+~ b_{n_y+1\da}) \Big\} \non \\
&& -~ i \lam_R ~\Big\{ \Big( \cos(\frac{\sqrt{3}k_x}{2}) ~+~ \sqrt{3} \sin 
(\frac{\sqrt{3}k_x}{2} ) \Big) a_{n_y+1\ua} ~-~ a_{n_y\ua}\Big\} ~+~ V_{n_y} ~
b_{n_y\da}. \label{eqns} \eea
\end{widetext}
Note that we have absorbed the lattice spacing $a$ into the definition of $k_x$
thereby making it a dimensionless quantity. By solving the above equations
numerically, we can obtain $E$ as a function of $k_x$.

In Fig.~\ref{lattice_disp}, we show $E$ versus $k_x$ for various cases. (In 
our calculations, we have taken $N_y = 200$ unit cells in the $y$-direction.
Hence the Hamiltonian is an $800 \times 800$ matrix due to the sublattice and 
spin degrees of freedom. We will also set $\ga = 1$ and the lattice spacing
$a=1$). Figs.~\ref{lattice_disp} (a), (b) and (c) show the energy spectrum 
when there
is no potential ($V_{n_y} = 0$), while Figs.~\ref{lattice_disp} (d), (e) and 
(f) show the spectrum in the presence of a single potential barrier which has 
a Gaussian shape. The width of the barrier is $4a$ and its peak value is $C = 
\ga/3$, where $\ga$ is the nearest-neighbor hopping amplitude. 
Figures~\ref{lattice_disp} (a) and (d) are for graphene without any SO 
couplings, i.e., $t_2 
= \lam_R = 0$. In Figs.~\ref{lattice_disp} (b) and (e), $t_2 = 0.02 ~\ga$ and 
$\lam_R = 0$, while in Figs.~\ref{lattice_disp} (c) and (f), $t_2 = \lam_R = 
0.02 ~\ga$. The blue shaded regions denote bulk states. The red dashed
lines show states which are localized along the barrier; their wave functions
decay exponentially as we go away from the barrier but are plane waves along 
the barrier. These one-dimensional states occur in a variety of systems
described by the Dirac equation, such as graphene~\cite{seshadri2} and
surfaces of three-dimensional topological insulators~\cite{deb12}.

We note that the modes localized along the barrier (shown by red dashed
lines in Figs.~\ref{lattice_disp} (d,e,f)) are not topologically protected. 
The modes in Figs.~\ref{lattice_disp} (d,f) are not topologically protected 
because the system is gapless and therefore in a non-topological phase on 
both sides of the barrier. The modes in Fig.~\ref{lattice_disp} (e) 
are not topologically protected because the system is in the same topological
phase on both sides of the barrier. % The lack of topological protection means,
%for instance, that if there is a non-magnetic impurity at some point on the 
%barrier, an electron can backscatter there with its momentum flipping from 
%$+ k_x$ to $- k_x$. Further, we find numerically that the number of states 
%localized along the barrier depends on the strength of the barrier. In fact, 
%the lower half of Figs. 4 (d,e,f) shows another set of such states as an 
%isolated solid blue line which merges with the bulk states close to the 
%Dirac points.

The states localized along the barrier have an interesting spin and sublattice 
structure. In Fig.~\ref{sublattice_spin}, we show the probabilities of $a\ua$,
$a\da$, $b\ua$ and $b\da$ as a function of the unit cell index $n_y$ for two 
states produced by a barrier of width $4a$ and peak value $\ga /3$; 
we have taken $t_2 = 0.02 ~\ga$ and $\lam_R = 0$. The two states are 
degenerate in energy, and
we see from the figure that the various probabilities in the two states
are related to each other by a simultaneous interchange of sublattice and 
spin. This symmetry follows from the observation that for $\lam_R =0$, 
Eqs.~\eqref{eqns} are invariant under the interchanges $a_{n_y \ua}
\leftrightarrow b_{-n_y \da}$ and $a_{n_y \da} \leftrightarrow b_{-n_y \ua}$, 
assuming that $V_{n_y} = V_{-n_y}$. (We note that this is the lattice version
of the symmetry of the continuum theory that was pointed out in 
Eq.~\eqref{sym3}).

\begin{widetext}
\begin{center}
\begin{figure}[H]
\begin{center}
%\subfigure[]{\includegraphics[width=5.2cm]{edited_Nx_200_Ny_200_t2_0_D_0_lr_0_C_0_w_4.eps}}
\subfigure[]{\includegraphics[width=5.2cm]{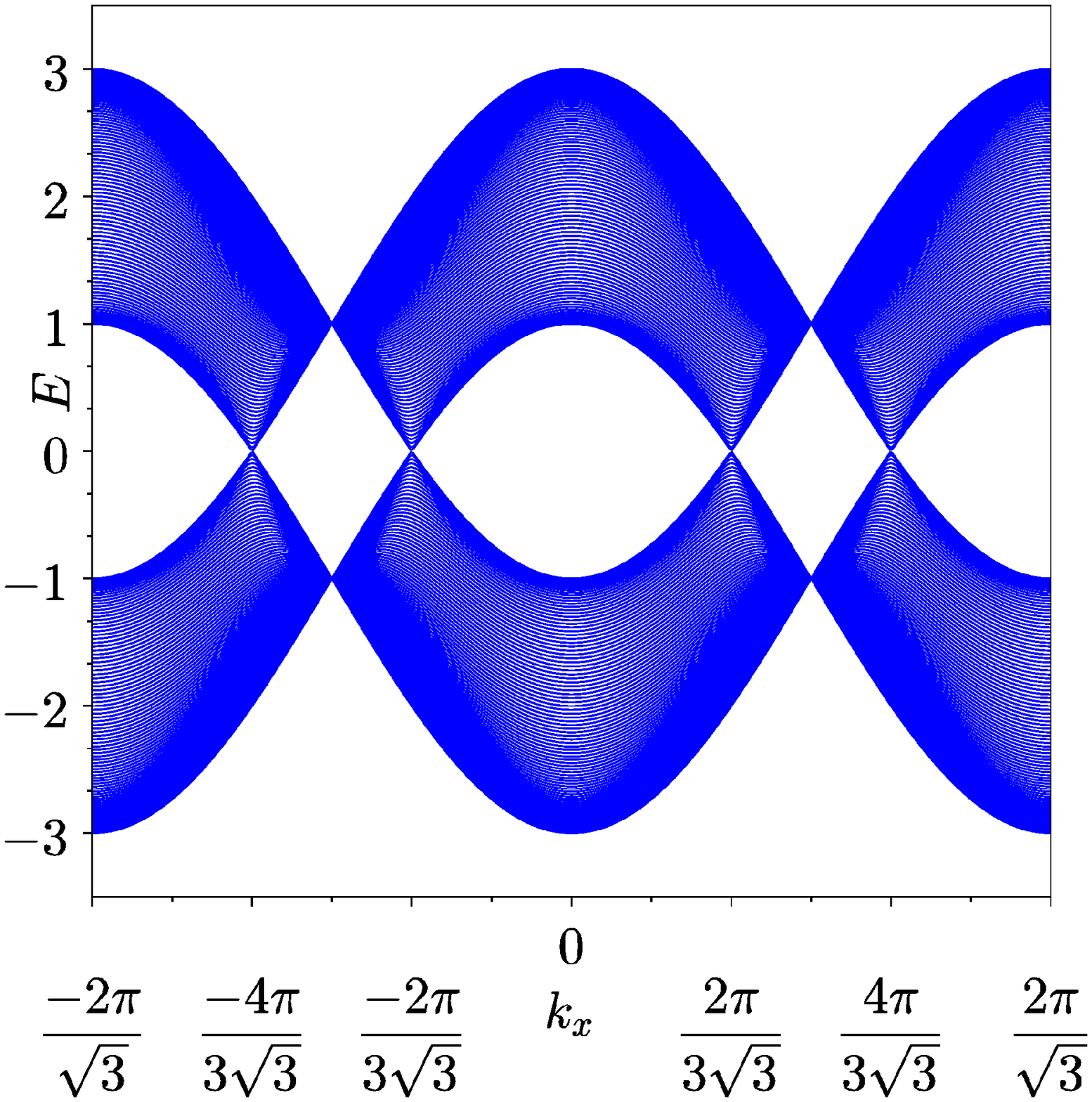}}
%\subfigure[]{\includegraphics[width=5.2cm]{edited_Nx_200_Ny_200_t2_0.02_D_3rt3t2_lr_0_C_0_w_4.eps}}
\subfigure[]{\includegraphics[width=5.2cm]{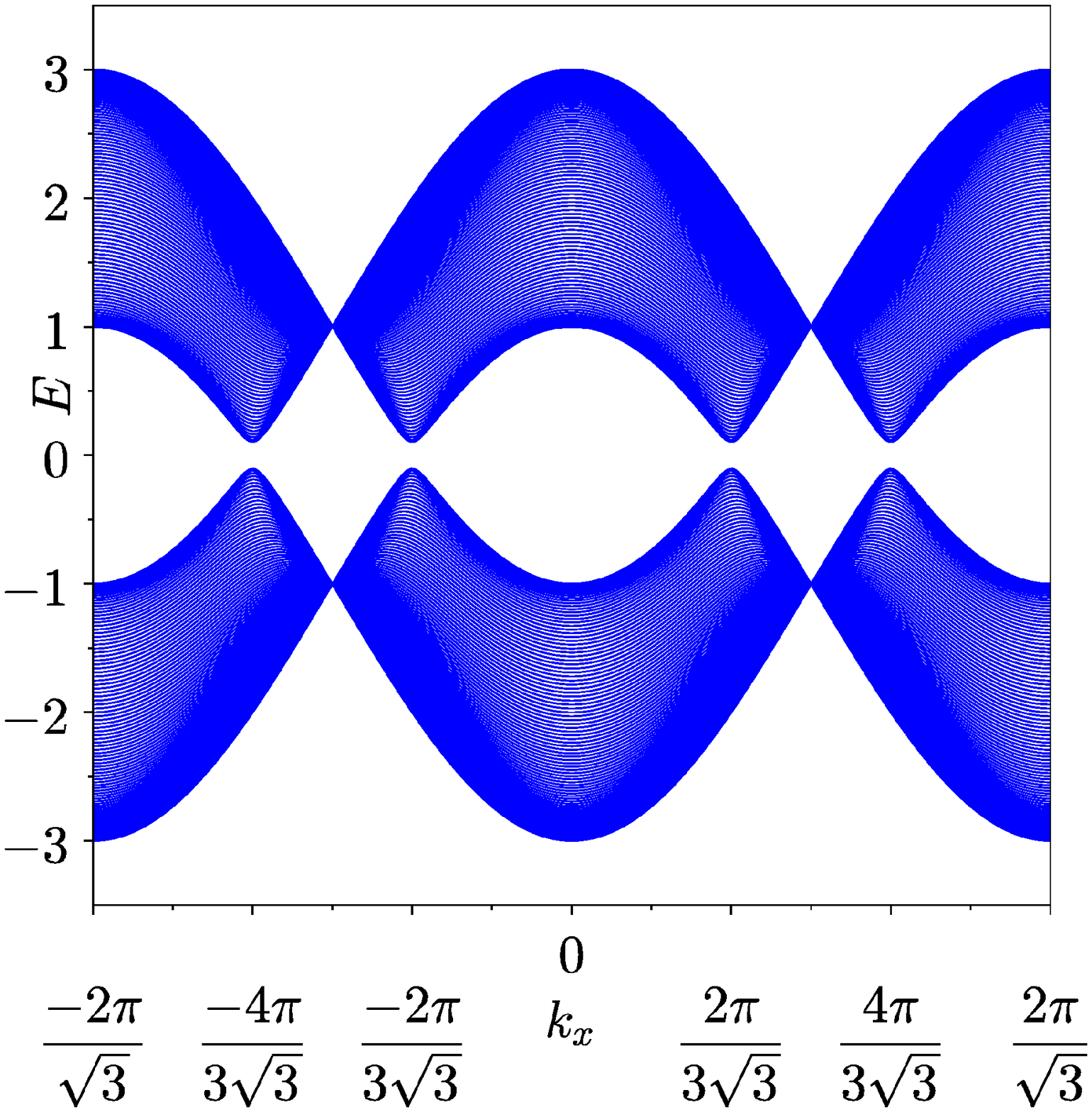}}
%\subfigure[]{\includegraphics[width=5.2cm]{edited_Nx_200_Ny_200_t2_0.02_D_3rt3t2_lr_D_C_0_w_4.eps}}
\subfigure[]{\includegraphics[width=5.2cm]{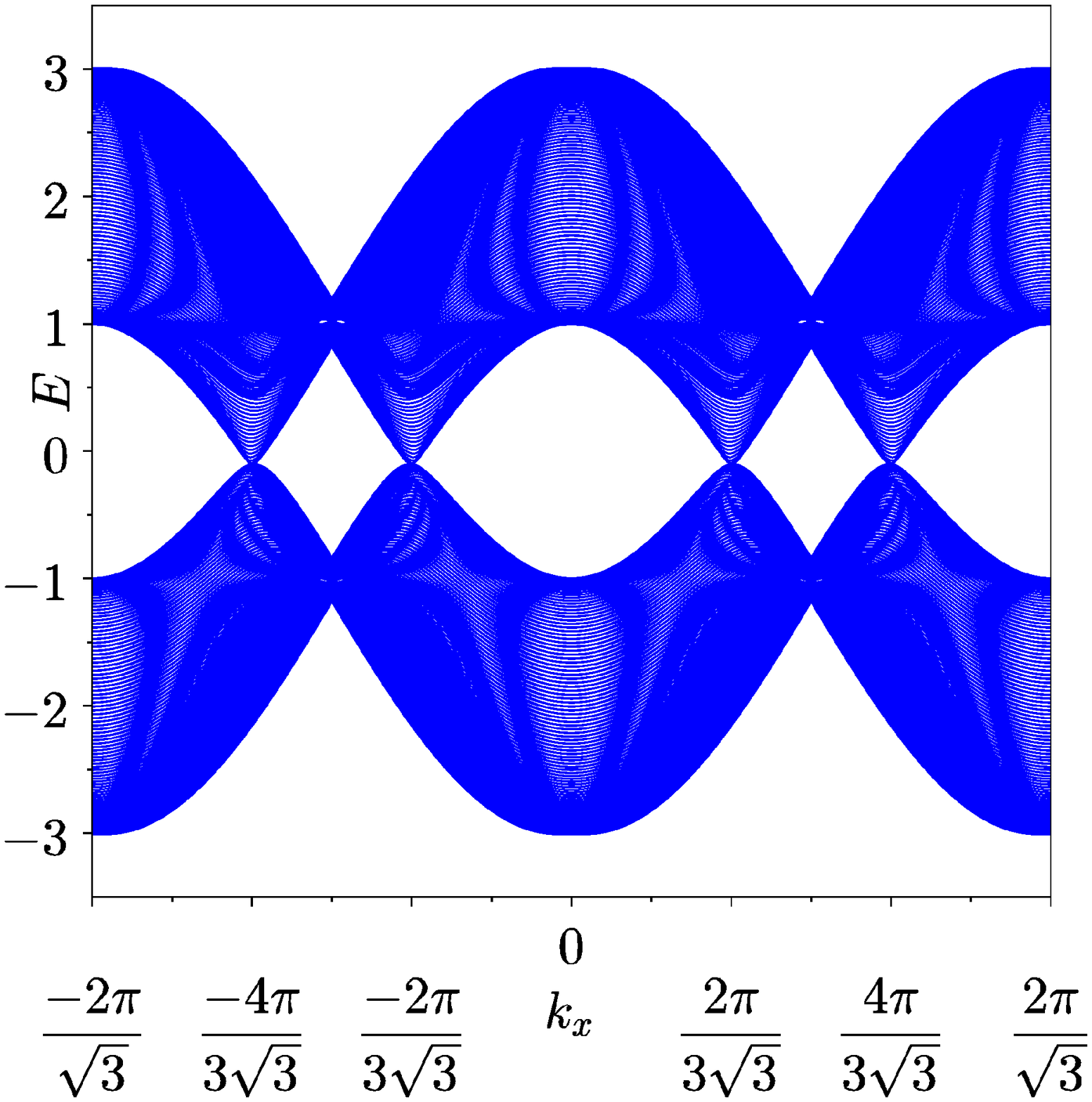}}
\\
%\subfigure[]{\includegraphics[width=5.2cm]{edited_zoom_Nx_200_Ny_200_t2_0_D_0_lr_0_C_t0by3_w_4.eps}}
\subfigure[]{\includegraphics[width=5.2cm]{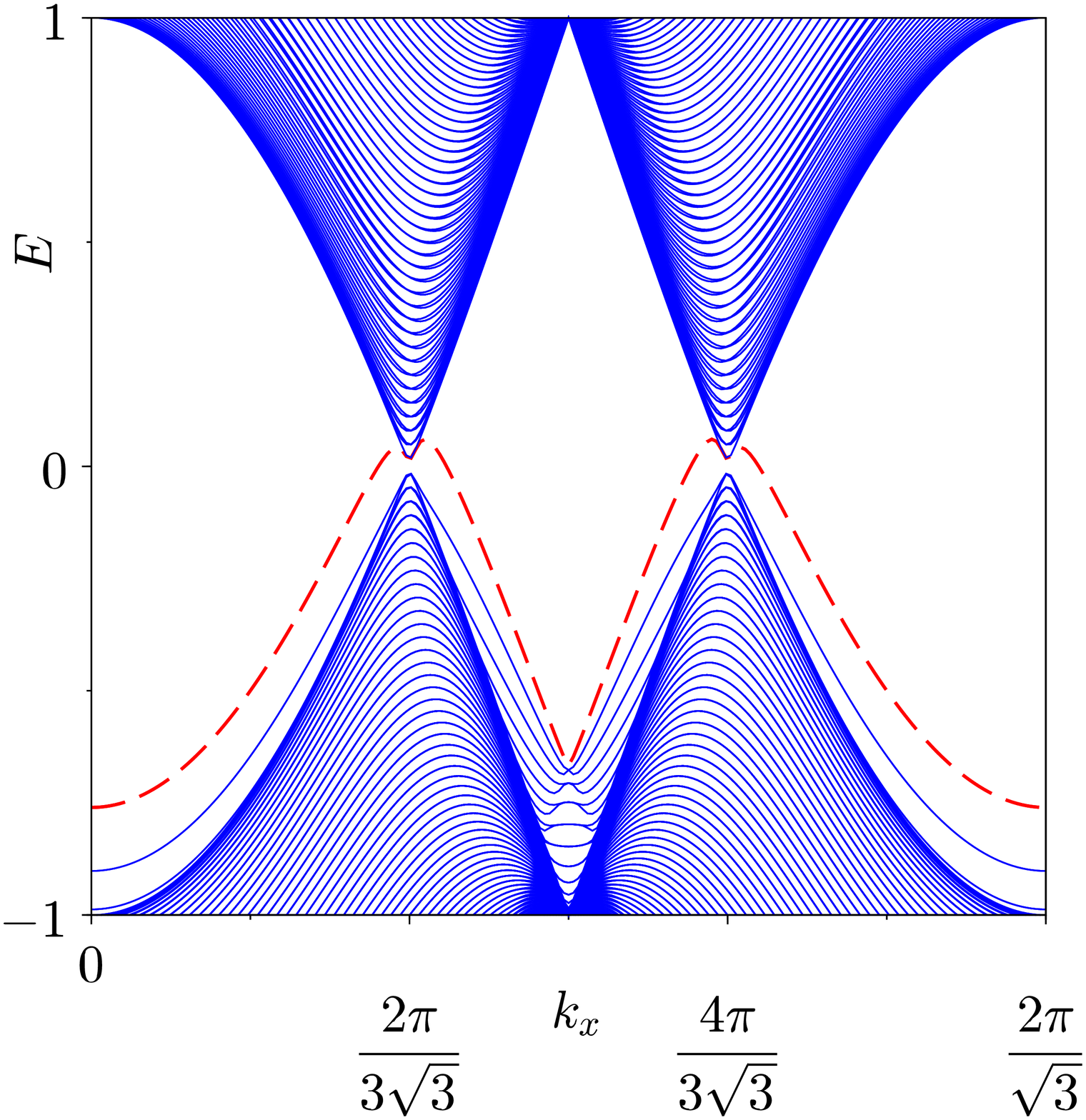}}
%\subfigure[]{\includegraphics[width=5.2cm]{edited_zoom_Nx_200_Ny_200_t2_0.02_D_3rt3t2_lr_0_C_t0by3_w_4.eps}}
\subfigure[]{\includegraphics[width=5.2cm]{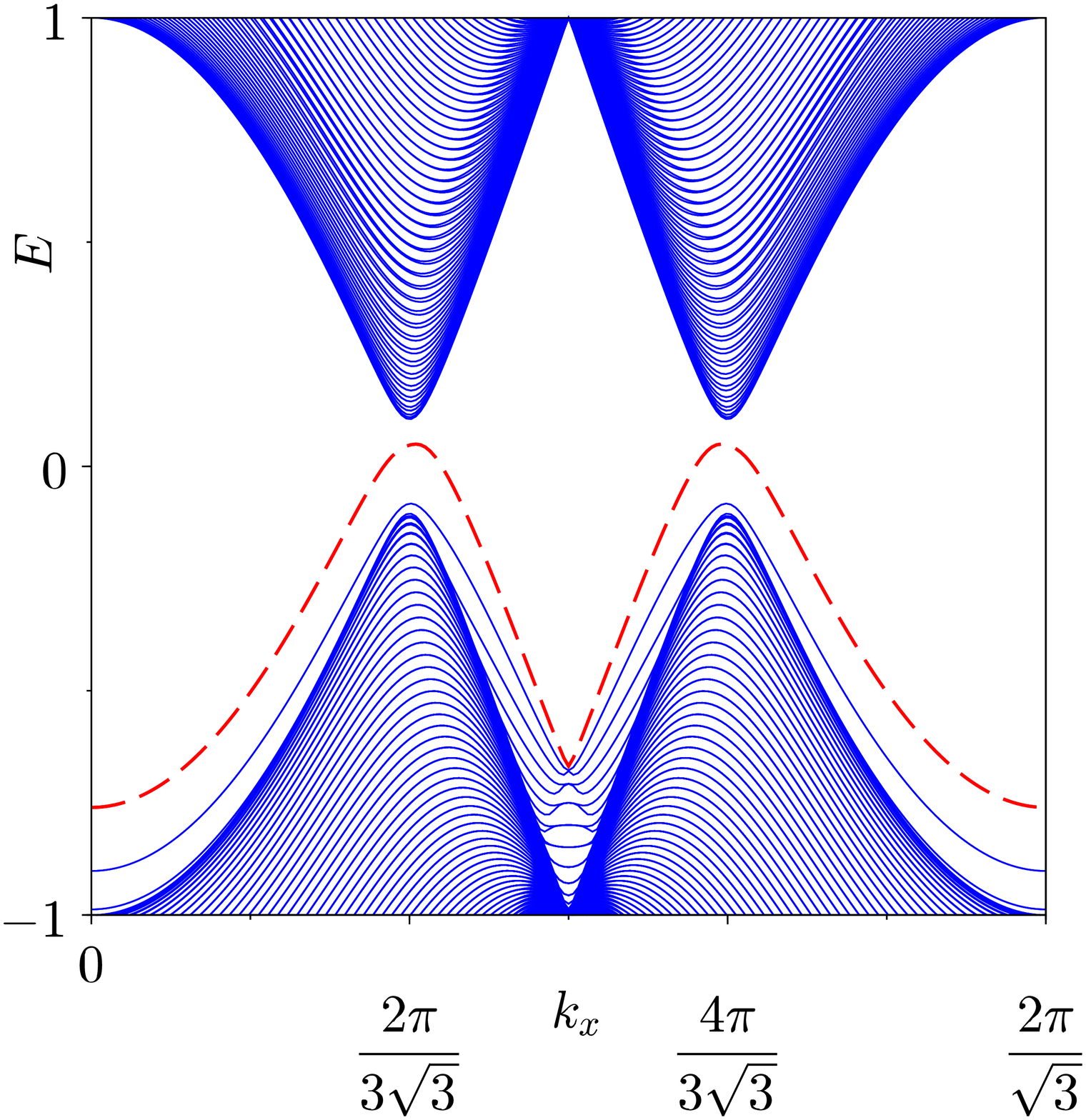}}
%\subfigure[]{\includegraphics[width=5.2cm]{edited_zoom_Nx_200_Ny_200_t2_0.02_D_3rt3t2_lr_D_C_t0by3_w_4.eps}}
\subfigure[]{\includegraphics[width=5.2cm]{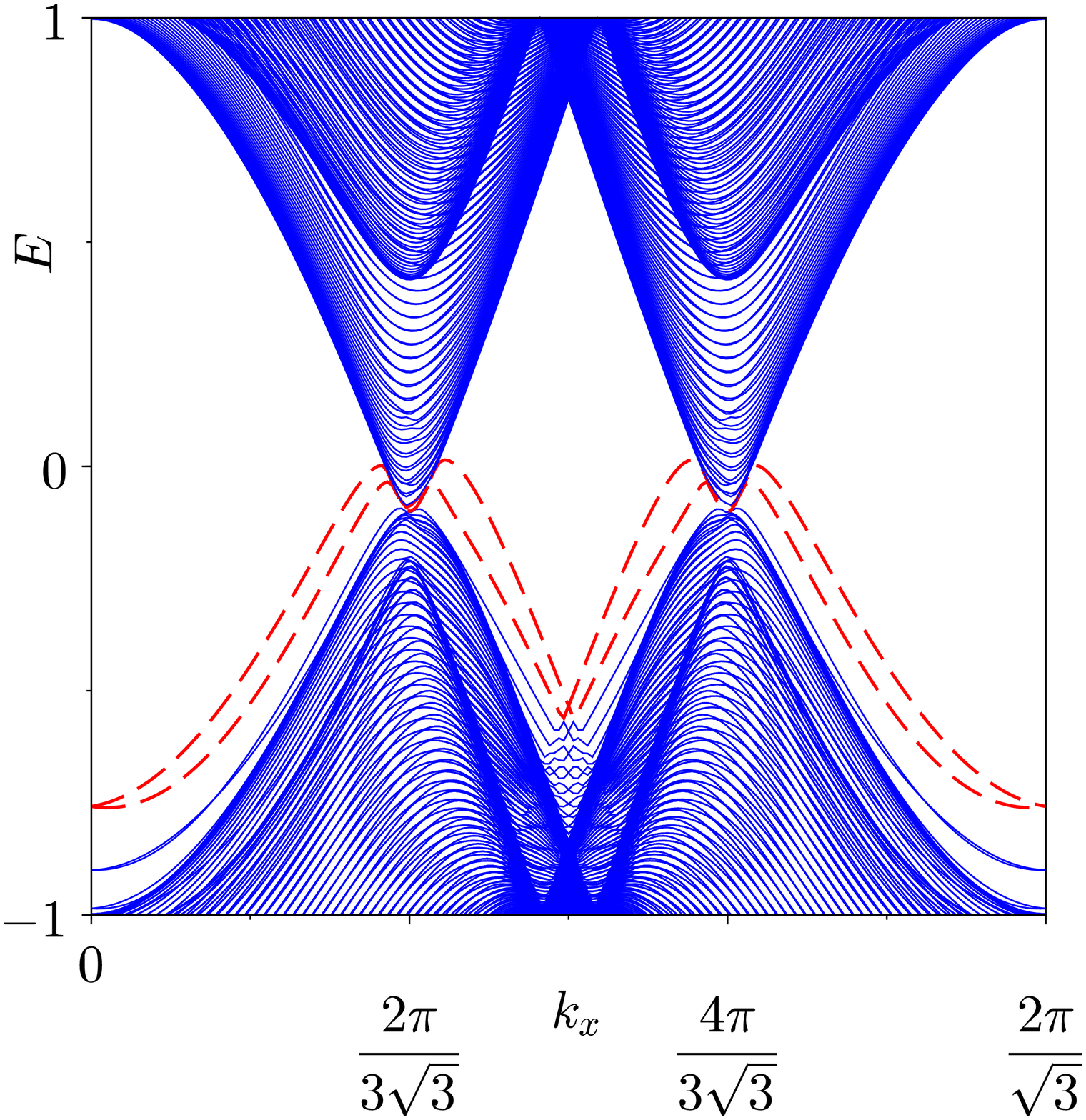}}
\end{center}
\caption{$E$ vs $k_x$ for various systems. ($k_x$ is in 
units of $1/a$ while $E$ is in units of $\ga$). Figures (a,b,c) and (d,e,f) 
show the spectrum without and with a single potential barrier respectively. 
The width of the barrier is $4a$ and its peak value is $C = \ga/3$. In 
Figs. (a) and (d), $\De_{KM} = \lam_R = 0$; in Figs. (b) and (e), $\De_{KM} 
= 0.1 ~\ga$ and $\lam_R 
= 0$; in Figs. (c) and (f), $\De_{KM} = \lam_R = 0.1 ~\ga$. The blue shaded 
regions denote bulk states while the red dashed lines in Figs. (d), (e) and 
(f) show states which are localized along the barrier. In Figs. (d) and (e),
the states localized along the barrier are doubly degenerate due to spin;
this degeneracy is broken in Fig. (f) due to the Rashba SO coupling.} 
\label{lattice_disp} 
\end{figure}
\end{center}
\end{widetext}

\begin{figure}[]
%\subfigure[]{\includegraphics[width=4.2cm]{PSI11_Nx_200_Ny_200_t2_0.02_D_3rt3t2_lr_0_C_t0by3_w_4.eps}}
\subfigure[]{\includegraphics[width=4.2cm]{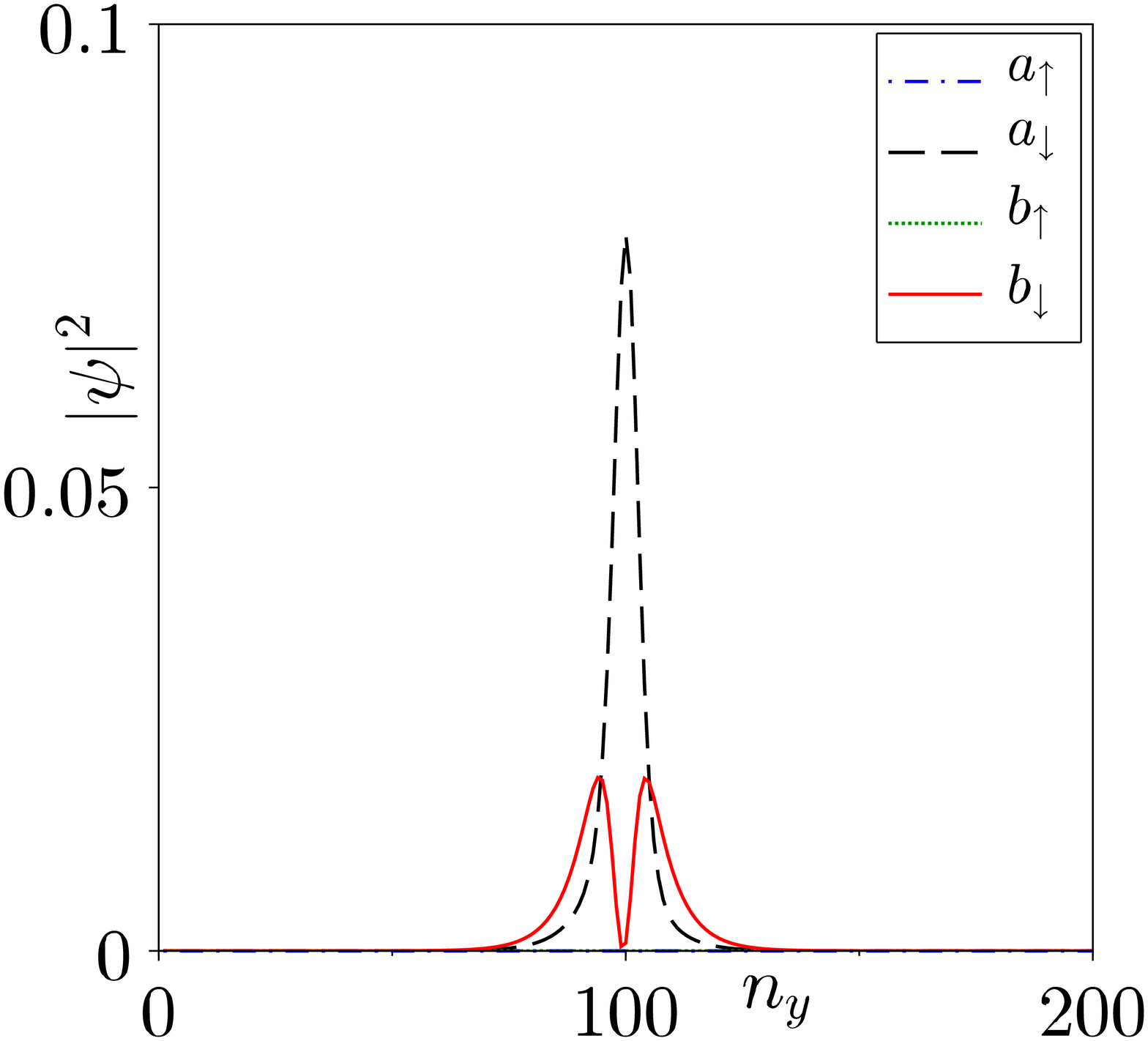}}
%\subfigure[]{\includegraphics[width=4.2cm]{PSI12_Nx_200_Ny_200_t2_0.02_D_3rt3t2_lr_0_C_t0by3_w_4.eps}}
\subfigure[]{\includegraphics[width=4.2cm]{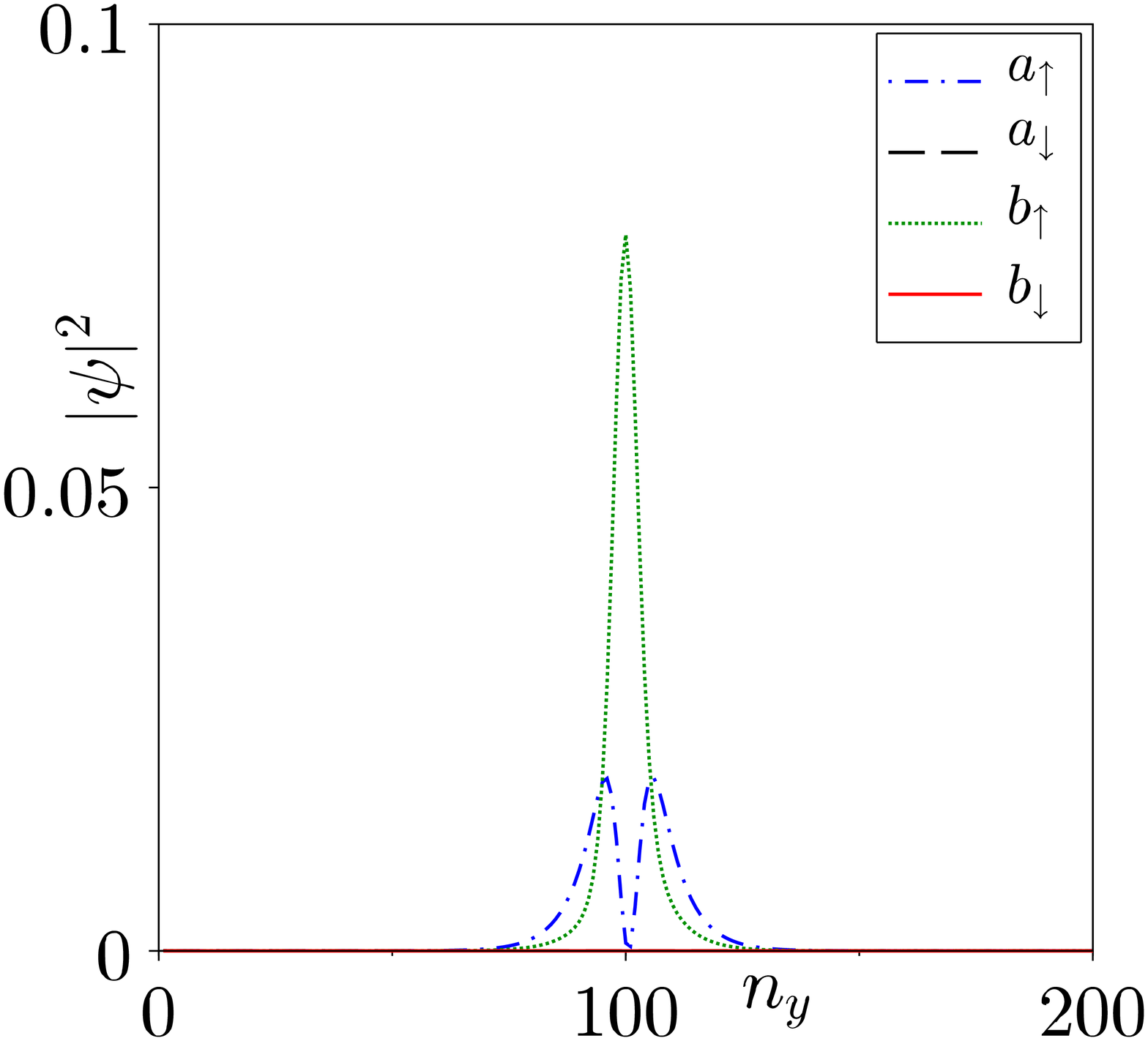}}
\caption{$|\psi|^2$ vs $n_y$ for two degenerate states localized along a 
barrier of width $4a$ and peak value $\ga/3$. The SO couplings are given by 
$\De_{KM} = 0.1 ~\ga$ and $\lam_R = 0$. The probabilities on sites $a \ua$, $a 
\da$, $b \ua$ and $b \da$ are shown by blue dot dash, black dashed, green dot 
and red solid lines respectively.} \label{sublattice_spin} \end{figure}

\section{Wave packet dynamics}

To numerically study the time evolution of a Gaussian wave packet on a 
honeycomb lattice, we take the rows of zigzag bonds to be parallel to the 
$x-$axis. Since our system has translational symmetry in the $x$ direction, 
$k_x$ is a good quantum number. However, periodic barriers parallel to 
the zigzag rows (Fig.~\ref{schematic}), break the translational 
invariance in the $y$ direction. We therefore consider a 
real space lattice with $N_y = 360$ unit cells in the $y$-direction. 
Hence, for every $k_x$, the Hamiltonian $H(k_x)$ is a $4N_y \times 4N_y$ 
matrix (accounting for the spin and sublattice degrees of freedom) and has 
periodic boundary condition in the $y$ direction. We denote the 
eigenvalues and eigenvectors of $H(k_x)$ by $E_{k_x}^{(n)}$ and 
$\phi_{k_x}^{(n)}$ respectively. 

We take the initial wave packet $\Psi(t=0)$ to be a Gaussian constructed such 
that it has a peak momentum $\vk_0$, peak position $\vec{r_0}$ and width 
$(W_x,W_y)$ of our choice. It is constructed out of the eigenvectors of 
the $4 \times 4$ lattice Hamiltonian $H(k_x,k_y)$ that we would get if both 
$k_x$ and $k_y$ were good quantum numbers; we choose the eigenvectors to lie 
within the positive energy band, $E_{k_x,k_y} \ge 0$. The width of the wave 
packet in momentum space is inversely proportional to the width 
in real space. Hence a Gaussian which is narrow in real space has a large
contribution from momenta far away from $\vk_0$, while a Gaussian which 
is wide in real space has contributions only from momenta which lie close to 
$\vk_0$. 

We incorporate periodic boundary conditions in the $x$ direction by taking 
$k_x$ in integer multiples of $2\pi /(N_x\sqrt{3}a)$; we have chosen
$N_x = 312$. We study the evolution of the wave packet by letting each of 
the momentum components $k_x$ evolve independently in time and then 
superposing them with suitable coefficients to form a Gaussian.

To summarize, let $\phi^{(n)}_{k_x}$ denote the $n$-th eigenvector of the
$4N_y \times 4N_y$ Hamiltonian $H(k_x)$, i.e.,
\beq H(k_x) \phi^{(n)}_{k_x} ~=~ E^{(n)}_{k_x} \phi^{(n)}_{k_x}. \eeq 
Next, $\phi^{(n)}_{k_x}$ consists of $N_y$ four-component
spinors each of which is labeled by the site index $n_y$; we denote these 
spinors by $| \phi^{(n)}_{k_x,n_y} \ra$. The four-component
spinor $|\Psi_{n_x,n_y} \ra = (a_{n_xn_y\ua}, a_{n_xn_y\da}, 
b_{n_xn_y\ua}, b_{n_xn_y\da})^T$ is then given by
\bea | \Psi_{n_x,n_y}(t) \ra &=& \sum_{k_x} e^{ik_x n_x} 
|{\tilde\Psi}_{k_x,n_y}(t) \ra, ~~~\text{and} \non \\
|{\tilde\Psi}_{k_x,n_y}(t) \ra &=& \sum_n e^{-i E^{(n)}_{k_x} t} | 
\phi^{(n)}_{k_x,n_y} \ra \la \phi^{(n)}_{k_x,n_y} | {\tilde\Psi}_{k_x,n_y}(0) 
\ra. \non \\
\eea
Using this formulation we study the propagation of a wave packet through 
the lattice. 

\subsection{Graphene with no spin-orbit couplings}

In this section, we study the time evolution of a Gaussian wave packet in
graphene without any SO couplings and without any potential barriers. We will 
show that there some special points in the Brillouin zone such that a wave
packet centered around those points does not spread significantly. 
However, wave packets centered around other momenta spread in time.

For graphene without any SO couplings, we can analytically find the following 
expressions for the energy and its derivatives; these are useful for 
understanding the time evolution of a wave packet.
\begin{scriptsize}
\bea E &=& \sqrt{3 + 2\cos{(\sqrt{3}k_x)} + 4\cos{(\frac{\sqrt{3}k_x}{2})}
\cos{(\frac{3k_y}{2})}}, \non \\
\frac{\pa E}{\pa k_x} &=& - \frac{\sqrt{3}}{E} \Big[\sin(\sqrt{3}k_x)+
\sin(\frac{\sqrt{3}k_x}{2})\cos(\frac{3k_y}{2})\Big], \non \\
\frac{\pa E}{\pa k_y}&=& -\frac{3}{E}\cos(\frac{\sqrt{3}k_x}{2})
\sin(\frac{3k_y}{2}), \non \\
\frac{\pa^2 E}{\pa k_x^2} &=& - \frac{1}{2E} \Big[3\cos(\sqrt{3}k_x) + 
2 (\frac{\pa E}{\pa k_x})^2 + 3\cos(\frac{\sqrt{3}k_x}{2}) 
\cos(\frac{3k_y}{2}) \Big], \non \\ 
\frac{\pa^2 E}{\pa k_x \pa k_y} &=& \frac{3}{E}\sin(\frac{3k_y}{2})
\Big[\frac{\sqrt{3}E}{2}\sin(\frac{\sqrt{3}k_x}{2}) +\cos
(\frac{\sqrt{3}k_x}{2})\frac{\pa E}{\pa k_x}\Big], \non \\
\frac{\pa^2 E}{\pa k_y^2} &=& - \frac{3}{E^2}\cos(\frac{\sqrt{3}k_x}{2}) 
\Big[\frac{3E}{2} \cos(\frac{3k_y}{2}) - \sin(\frac{3k_y}{2})
\frac{\pa E}{\pa k_y} \Big]. \eea
\end{scriptsize}
While the first derivatives represent the group velocities in the $x$ and $y$ 
directions, the second derivatives give an estimate of the rate at which the 
width of the wave packet changes. This can be qualitatively 
understood as follows. Given a 
wave packet centered around $(k_x,k_y)$, the group velocity is $\vec{v}_g = 
(\pa E /\pa k_x, \pa E /\pa k_y)$. However, since the wave packet has momentum
components lying in a finite range $(k_x \pm \de k_x, k_y \pm \de k_y)$, the 
group velocity itself will have a spread given by $\de k_x \pa \vec{v}_g /\pa 
k_x$ and $\de k_y \pa \vec{v}_g/\pa k_y$ which involve the second derivatives 
of $E$. The spread in the group velocity determines the rate at which the 
width of the wave packet changes. More quantitatively, let us consider a
wave packet moving in one dimension which, at time $t=0$, is centered at 
$k_0$ and $x_0$ in momentum and real space and has width $W_x$ in real space. 
The momentum component of such an object is given by
\beq {\tilde \psi} (k_x,0) ~\sim~ \exp [i k_x (x - x_0) -W_x^2 ~(k_x - k_0)^2].
\eeq
When this is evolved in time with energy 
\beq E ~=~ E_0 ~+~ (k_x - k_0) E'_0 ~+~ \frac{1}{2} (k_x - k_0)^2 E''_0, \eeq
where $E'_0$ and $E''_0$ denote the first and second derivatives of $E$ with 
respect to $k_x$ evaluated at $k_x = k_0$, we obtain
\bea {\tilde \psi} (k_x,t) &\sim& \exp \Big[ ik_x (x- x_0) - i E_0 t - i 
(k_x - k_0) E'_0 t \non \\
&& ~~~~~- (W_x^2 + i E''_0 t/2) (k_x - k_0)^2 \Big]. \non \\
\eea
Fourier transforming this and taking the modulus squared gives the probability
density in real space
\beq |\psi (x,t) |^2 ~\sim~ \exp \left[ - \frac{(x-x_0 - E'_0 t)^2}{2 (W_x^2 +
(E''_0 t /(2W_x))^2)} \right]. \eeq
This shows that the width in real space evolves as 
\beq W (t) ~=~ \sqrt{W_x^2 + \left( \frac{E''_0 t}{2 W_x} \right)^2}. \eeq
Thus at long times (when $t \gg 2 W_x^2 /E''_0$), the width increases linearly
with time at a rate given by $(1/2W_x) (\pa^2 E/\pa k_x^2)_{k_x = k_0}$.

While it is not unusual to have points in a one-dimensional Brillouin zone 
where the second derivative of $E$ with respect to the momentum vanishes, 
it is not common to find two-dimensional models in which all the three second 
derivatives of $E$ (namely, $\pa^2 E/\pa k_x^2$, $\pa^2 E/\pa k_y^2$ and
$\pa^2 E/\pa k_x \pa k_y$) vanish at certain points. For instance, all three 
second derivatives do not vanish simultaneously even for the Dirac dispersion
$E = v_F \sqrt{k_x^2 + k_y^2}$. Thus graphene is a rare example of a 
system with a number of no-spreading points where all the second derivatives
vanish.

\begin{figure}[H]
\begin{center}
%\subfigure{\ig[width=6.5cm]{level_curves.eps}}
\subfigure{\ig[width=6.5cm]{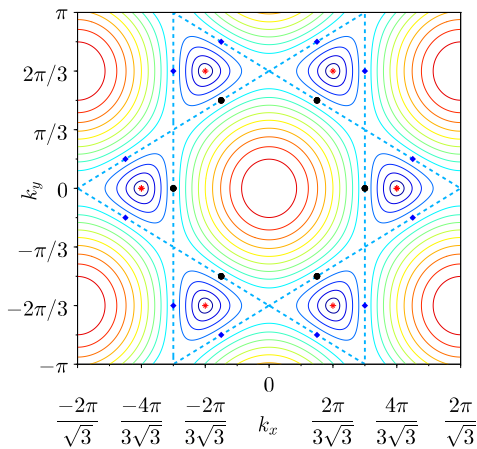}}
\end{center}
\caption{Level curves for the positive energy band of graphene without any SO 
couplings. ($k_x, ~k_y$ are in units of $1/a$). The figure shows an area equal
to three Brillouin zones and six Dirac points (red stars). The blue dashed 
line is the locus of points with $E= \ga$. The six black dots at $(k_x,k_y) 
= (\pm \pi/(\sqrt{3}a), 0)$ and $(\pm \pi/(2\sqrt{3}a), \pm \pi/(2a))$ are 
the no-spreading points in the first Brillouin zone where the second 
derivatives of $E(k_x,k_y)$ vanish. The blue diamonds denote the corresponding
points in the neighboring Brillouin zones and are related to the former set 
of points by reciprocal lattice vectors.} \label{beautiful_star}
\end{figure}

Figure~\ref{beautiful_star} shows the level curves for the positive
energy (conduction) band of graphene. The Dirac 
points, where $E=0$, lie at $(\pm 4\pi/(3 \sqrt{3}a),0)$, $(\pm 2\pi/(3 
\sqrt{3}a),\pm 2\pi/(3a))$ and are shown as red stars. The figure also shows 
six points where the second derivatives of $E$ vanish. Within the first 
Brillouin zone these are located at $(\pm \pi/(\sqrt{3}a),0)$ and $(\pm \pi/
(2\sqrt{3}a),\pm \pi/(2a))$ and are marked as black dots. (The blue diamond 
marks denote the 
corresponding points in the neighboring Brillouin zones and are related
to the former set of points by the reciprocal lattice vectors). A wave packet
whose momentum components are centered around any of these points should move
through the lattice without any significant spreading. We will therefore call 
these the ``no-spreading points". The distances of these points from the 
center of the Brillouin zone is $3/4$ of the distances of the Dirac points. 
Interestingly, the no-spreading points lie on the lines with $E=\ga$ which is 
the energy at which the density of states has a Van Hove 
singularity~\cite{neto09}.
[We note that the existence of no-spreading points is specific to a lattice 
model. A continuum model of either massless or massive Dirac fermions (with 
$E = \pm v_F | \vec k|$ or $\pm \sqrt{v_F^2 \vk^2 + M^2}$) does not have any 
points in momentum space where all the second derivatives of $E$ vanish].

In Figs.~\ref{TE_00} and \ref{TE_nospread}, we show the time evolution of
wave packets centered at two different points in momentum $(k_x,k_y)$, namely,
the origin $(0,0)$ and a no-spreading point $(-\pi/(2\sqrt{3}a),-\pi/(2a))$;
at $t=0$, the wave packets are taken to have real space width $W_x = W_y = 8$
in units of the lattice spacing $a$. At $(k_x,k_y) = (0,0)$, the energy 
spectrum is 
flat; hence the group velocity $\vec{v}_g$ is zero along both $x$ and $y$ 
directions. Thus a wave packet with a peak momentum at $(0,0)$ which is 
centered around a point in real space continues to be centered around the 
same point as it evolves in time. However, it spreads uniformly in all 
directions as $\pa^2 E /\pa k_x^2$ and $\pa^2 E /\pa k_x^2$ are non-zero and 
equal, while $\pa^2 E/ \pa k_x \pa k_y = 0$. The behavior of such a 
wave packet is shown in Fig.~\ref{TE_00}. We find that the wave packet 
spreads out isotropically; the spread 
increases linearly with time at long times (Fig.~\ref{TE_00_spread}).
In contrast to this, at $(k_x,k_y) = (-\pi/(2\sqrt{3}a),-\pi/(2a))$, the 
group velocity is non-zero but the second derivatives of $E$ are zero. 
As Fig.~\ref{TE_nospread} shows, a wave packet centered around this
momentum moves but does not spread.

\begin{widetext}
\begin{center}
\begin{figure}[H]
\begin{center}
\subfigure[~Wave packet with peak momentum $\vk_0 = (0,0).$]
%{\ig[height=3.8cm]{edited_dyn_pristine_00.eps}
{\ig[height=3.8cm]{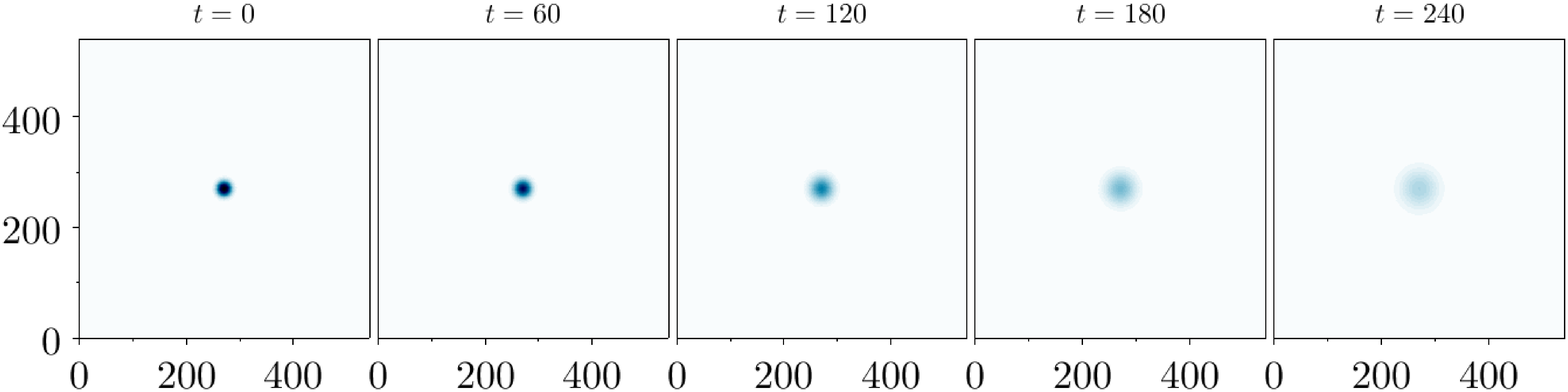}
%\ig[height=3.8cm]{edited_col_pristine_00.eps} \label{TE_00}}
\ig[height=3.8cm]{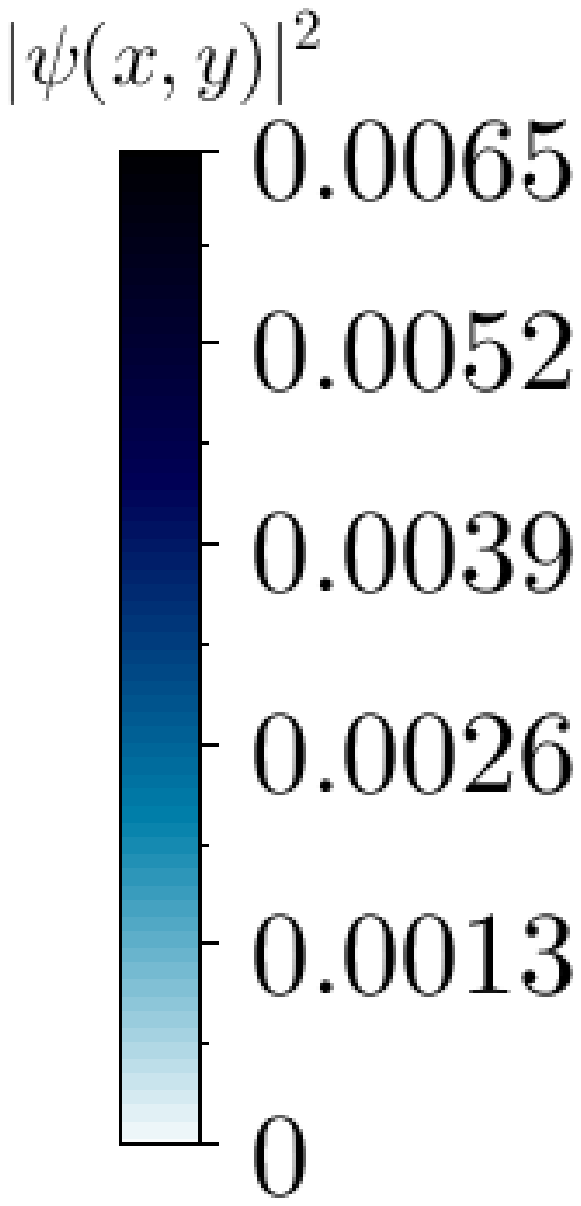} \label{TE_00}}
\subfigure[~Wave packet with peak momentum $\vk_0 = (-\pi/(2\sqrt{3}a),
%-\pi/(2a)).$] {\ig[height=3.8cm]{edited_dyn_pristine_nospread.eps}
-\pi/(2a)).$] {\ig[height=3.8cm]{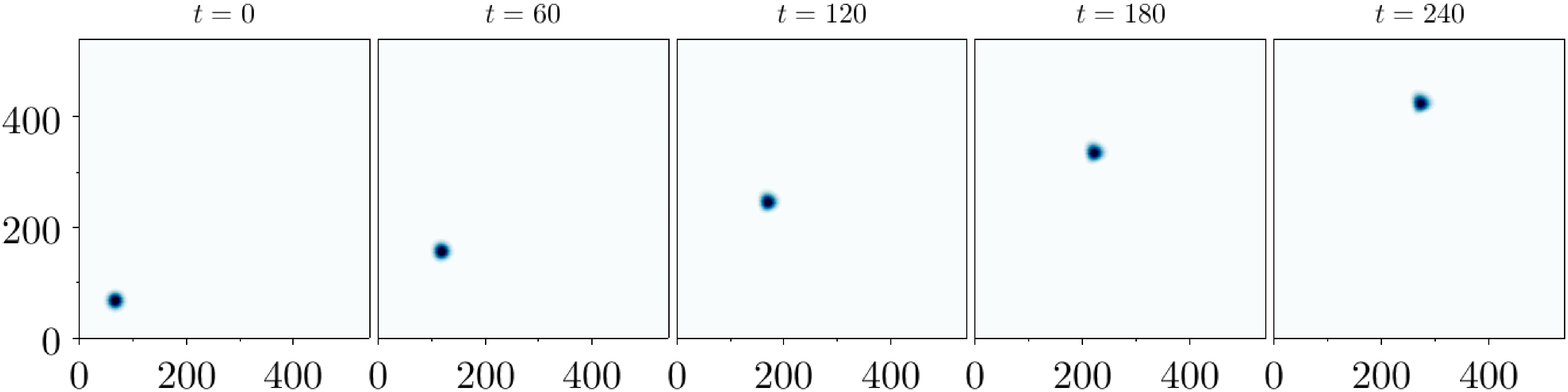}
%\ig[height=3.8cm]{edited_col_pristine_nospread.eps}\label{TE_nospread}}
\ig[height=3.8cm]{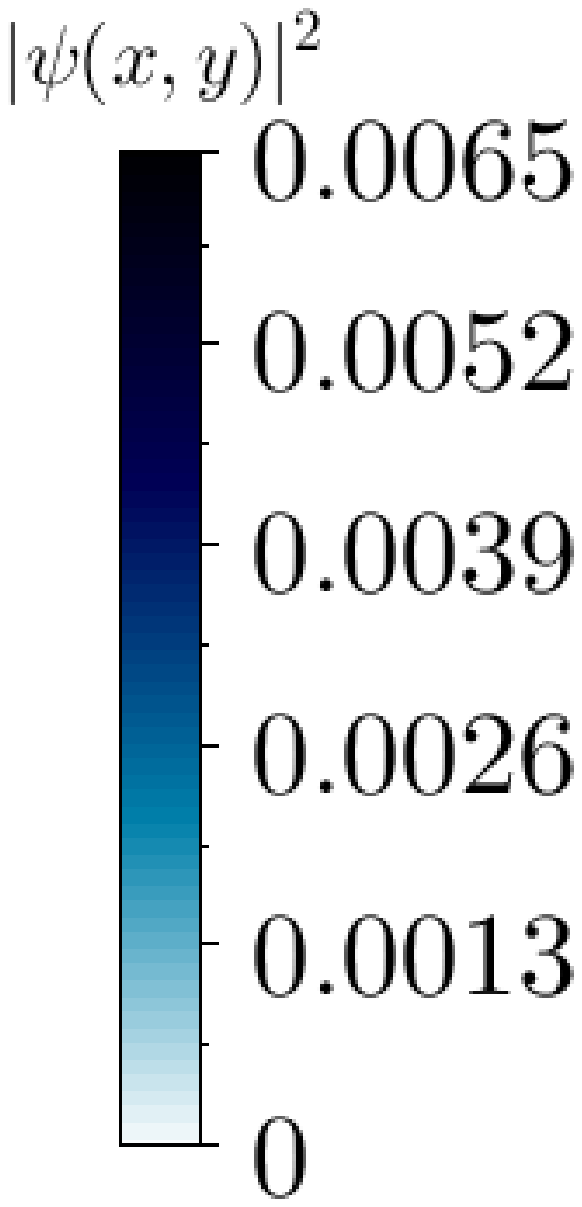}\label{TE_nospread}}
\end{center}
\caption{Evolution of a Gaussian wave packet created in graphene without any
SO couplings at 
$t=0$ with width $W_x = 8a$ and $W_y = 8a$. The $x$ and $y$ coordinates 
(horizontal and vertical directions respectively) are in units of
$a$ while $t$ is in units of $\hbar /\ga$. In (a), since the group 
velocity $\vec{v}_g (k_x = 0, k_y = 0) = \nabla_{\vk}{E}|_{(0,0)} = (0,0)$, 
the wave packet does not move. But it spreads as $\pa^2 E/\pa k_x^2$ and 
$\pa^2 E/ \pa k_y^2$ are non-zero. In (b), $\vec{v}_g (k_x = -\pi/
(2\sqrt{3}a), k_y = -\pi/(2a)) \neq 0$ but the second derivatives of $E$ 
vanish; this momentum is one of the no-spreading points shown in 
Fig.~\ref{beautiful_star}. Hence this wave packet moves but does not spread.}
\end{figure}
\end{center}
\end{widetext}

\begin{center} 
\begin{figure}[H]
\begin{center}
%\ig[height=6.5cm]{spread_pristine_00.eps}
\ig[height=6.5cm]{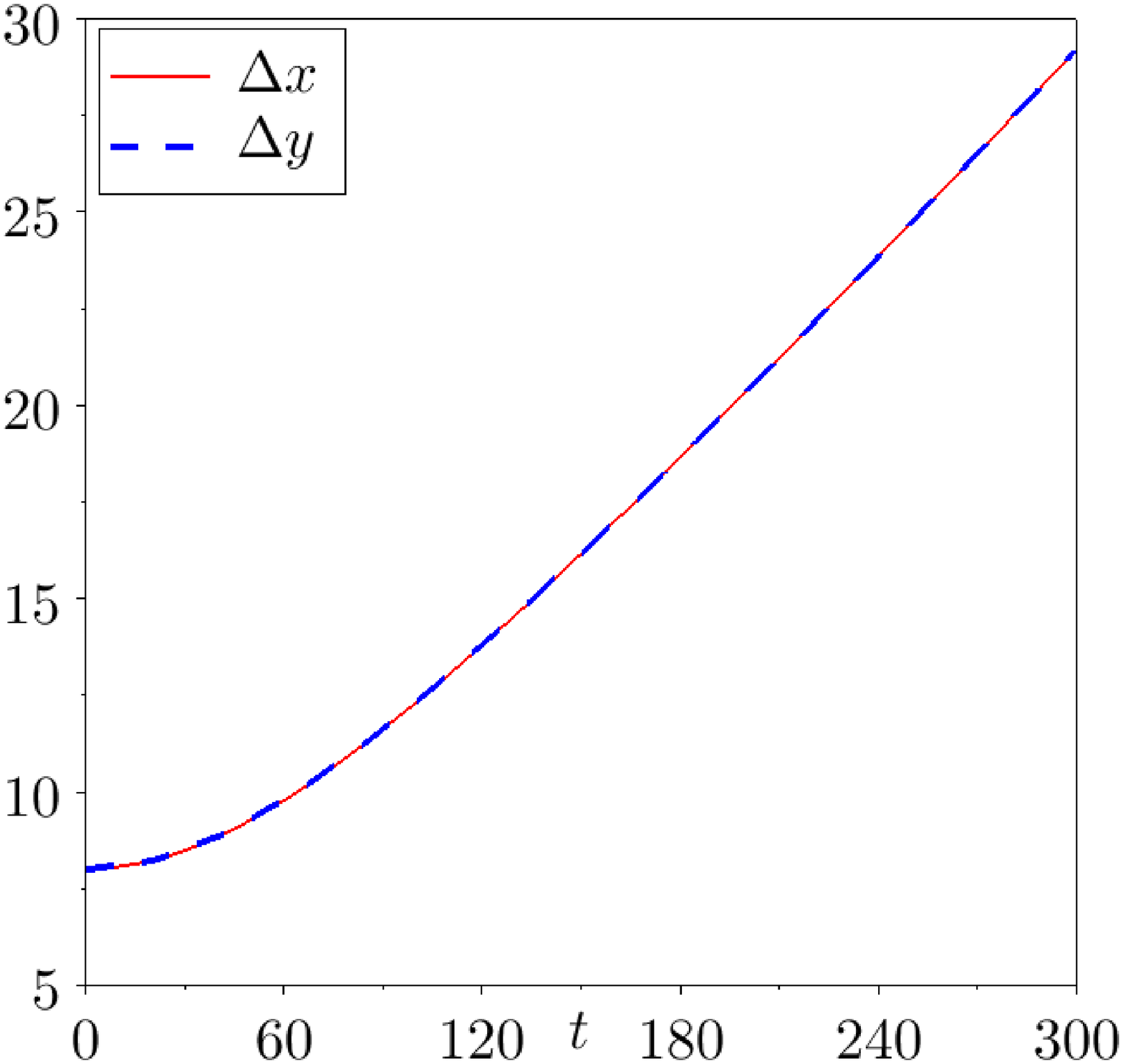}
\end{center}
\caption{Spreads $\De x$ and $\De y$ in the $x$ and $y$ directions of the wave
packet with peak momentum $\vk_0 = (0,0)$ as a function of time. (Both $\De x$
and $\De y$ are in units of $a$ while $t$ is in units of $\hbar /\ga$). The 
time evolution of the wave packet is shown in Fig.~\ref{TE_00}.} 
\label{TE_00_spread} 
\end{figure}
\end{center}

\subsection{Periodic potential barriers}

Next we look at the behavior of a wave packet when periodic potential barriers 
of the form shown in Fig.~\ref{schematic} are present. We first consider a wave
packet whose momentum components are centered around one of the no-spreading 
points $\vk_0 = (-\pi/(2\sqrt{3}a),-\pi/(2a))$. Since each barrier is quite 
high ($C=\ga$) and the wave packet has no components close to any of the Dirac 
points, there is almost no Klein tunneling and the reflection probability is 
close to 1. Hence the wave packet just reflects back and forth and stays 
between two successive barriers. This is shown in 
Fig.~\ref{TE_barrier_nospread}. The wave packet becomes narrower at the 
instant when it hits a barrier and is about to reflect back; this is visible 
in the second and fourth panels of Fig.~\ref{TE_barrier_nospread}. However, 
the width of the wave packet does not change when it is far from the barriers.

In contrast, when a wave packet is built with momenta 
centered around $\vk_0 = (4\pi/(3\sqrt{3}a), \pi/(5a))$ which lies close to a
Dirac point, we see in Fig.~\ref{TE_barrier_Diracpt} that it Klein tunnels
through the barriers, each of height $C = (2/3) \ga$. Since a narrower wave 
packet spreads faster, we have chosen a larger width $W_x = W_y = 16a$ in 
order to clearly show the Klein tunneling near the Dirac point. Note that we 
have not taken the wave packet to be centered around a Dirac point exactly 
since the group velocity is not well defined at those points.

\begin{widetext}
\begin{center}
\begin{figure}[H]
\begin{center}
\subfigure[~Wave packet with peak momentum $\vk_0 = (-\pi/(2\sqrt{3}a),-
\pi/(2a))$.]
%{\ig[height=3.5cm]{edited_dyn_pristine_nospread_C1.eps}
{\ig[height=3.5cm]{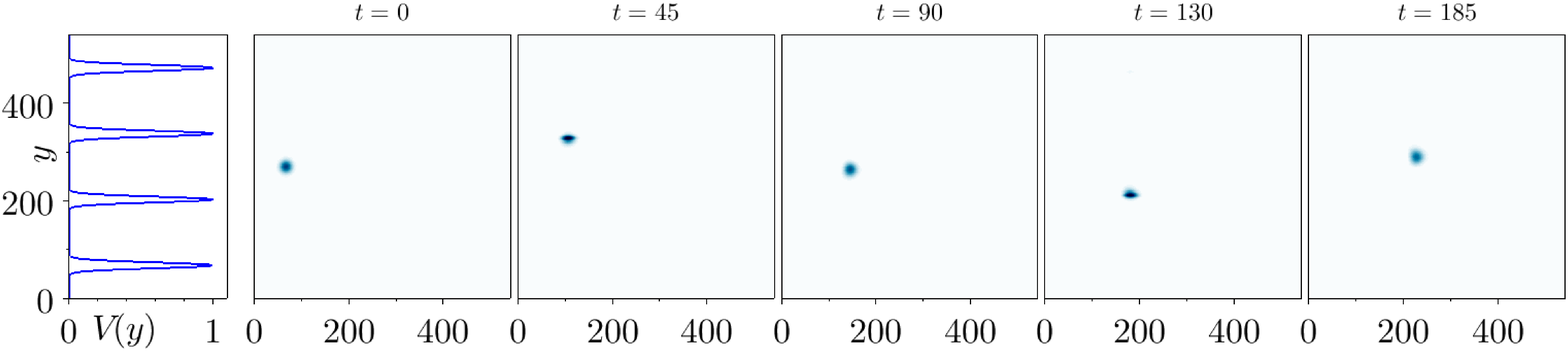}
%\ig[height=3.5cm]{edited_col_pristine_nospread_C1.eps}
\ig[height=3.5cm]{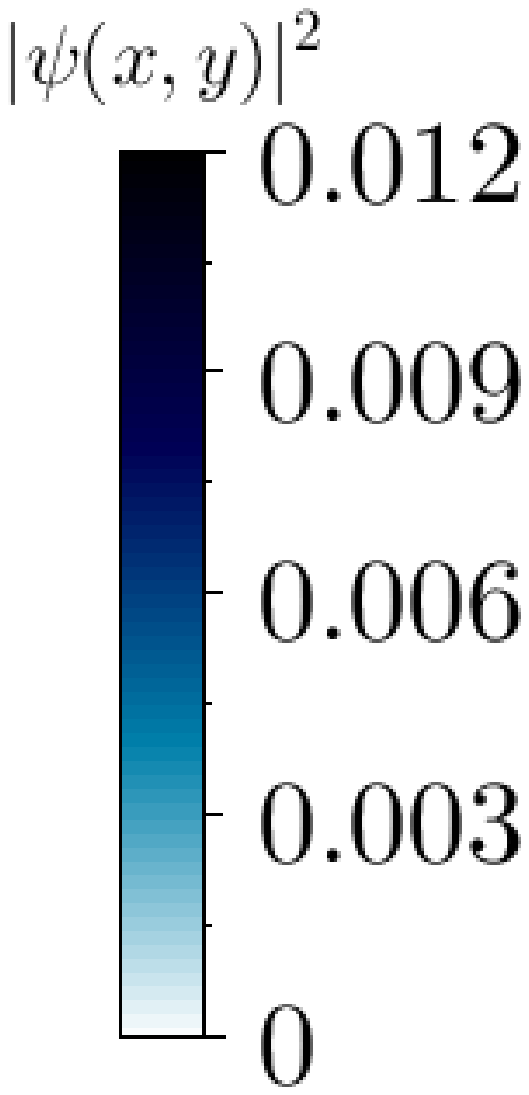}
\label{TE_barrier_nospread}}
\subfigure[~Wave packet with peak momentum $\vk_0 = (4\pi/(3\sqrt{3}a),\pi/
(5a))$.]
%{\ig[height=3.5cm]{edited_dyn_pristine_Dirac_C2by3.eps}
{\ig[height=3.5cm]{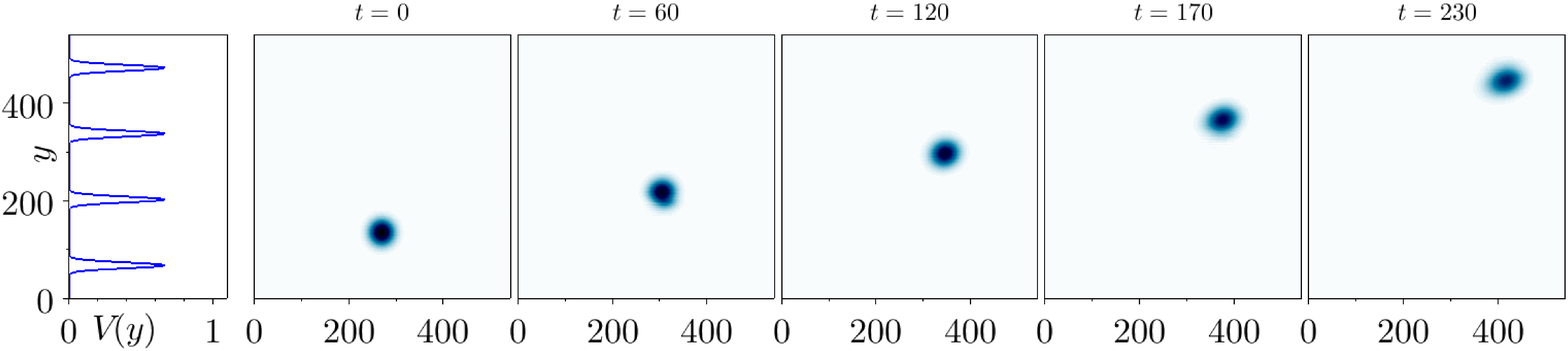}
%\ig[height=3.5cm]{edited_col_pristine_Dirac_C2by3.eps}
\ig[height=3.5cm]{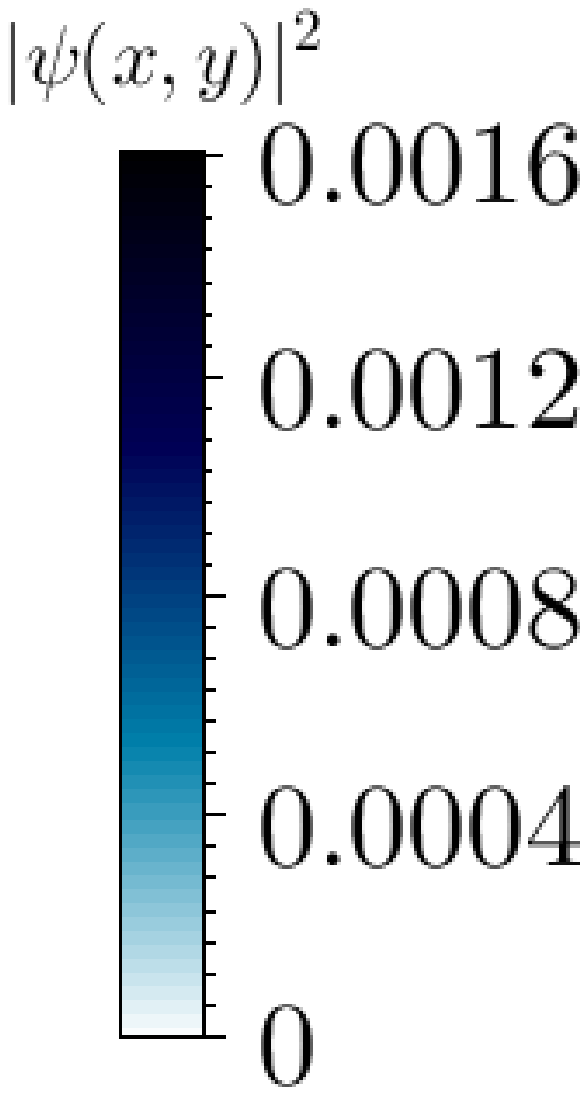}
\label{TE_barrier_Diracpt}}
\end{center}
\caption{Evolution of a Gaussian wave packet in graphene without any SO
couplings in the
presence of equally spaced barriers of strength $C = \ga$; the spacing between
the barriers is $d=135a$, i.e., 90 unit cells. The $x$ and $y$ coordinates
(horizontal and vertical directions respectively) are in units of $a$ while 
$t$ is in units of $\hbar /\ga$. The leftmost panels 
show the positions of the barriers. In (a) the wave packet has width $W_x = 
W_y = 8a$ at $t=0$. Due to the large barrier height, it almost completely 
reflects back and forth between two barriers. However it does not spread as 
it is centered around the no-spreading point $(-\pi/(2\sqrt{3}a),-\pi/(2a))$ 
in momentum space. In (b) the wave packet has initial width $W_x = W_y = 16a$ 
and is centered around $(4\pi/(3\sqrt{3}a), \pi/(5a))$ which is close to a 
Dirac point. It almost completely Klein tunnels through the barriers, each of 
strength $C=(2/3) \ga$.}
\end{figure}
\end{center}
%\end{widetext}
\vspace*{.4cm}

%\begin{widetext}
\begin{center}
\begin{figure}[H]
\begin{center}
\subfigure[~Gapped mode]
%{\ig[height=3.5cm]{edited_dyn_KMR_Dirac_C1_gapped.eps}
{\ig[height=3.5cm]{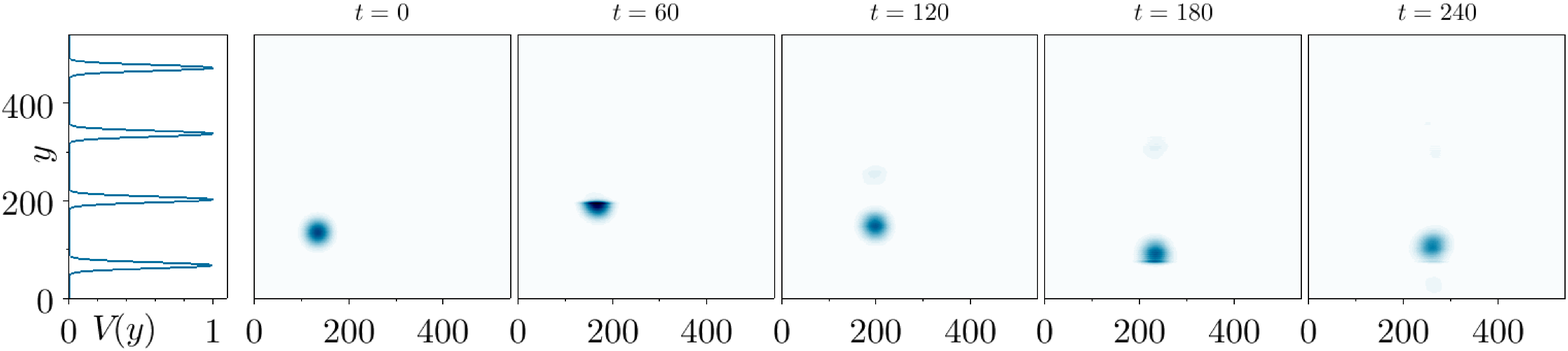}
\label{TE_barrier_Dirac_gapped}
%\ig[height=3.5cm]{edited_col_KMR_Dirac_C1_gapped.eps}}
\ig[height=3.5cm]{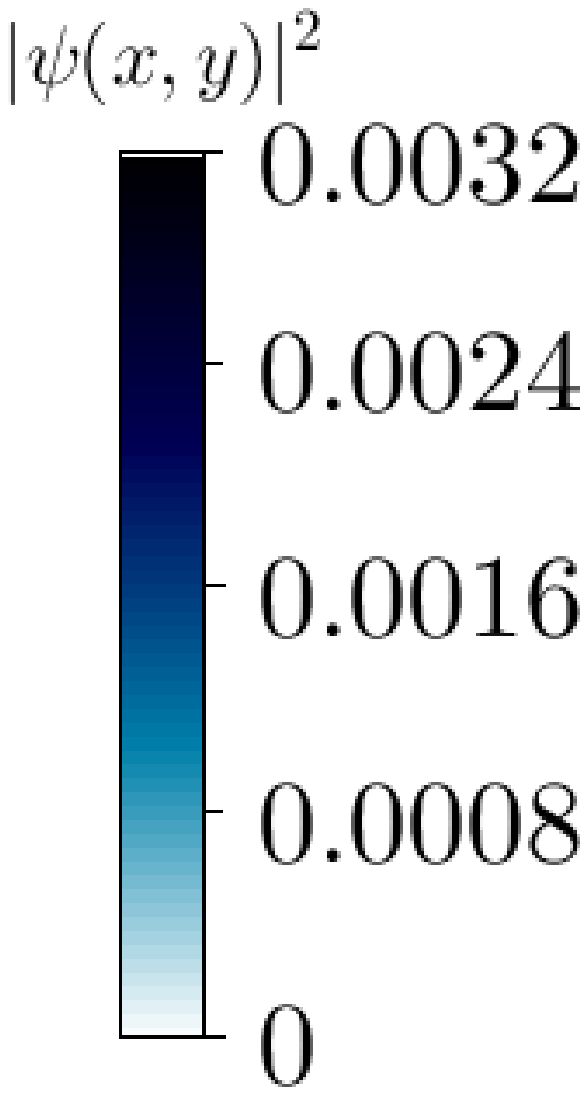}}
\subfigure[~Gapless mode]
%{\ig[height=3.5cm]{edited_dyn_KMR_Dirac_C1_gapless.eps}
{\ig[height=3.5cm]{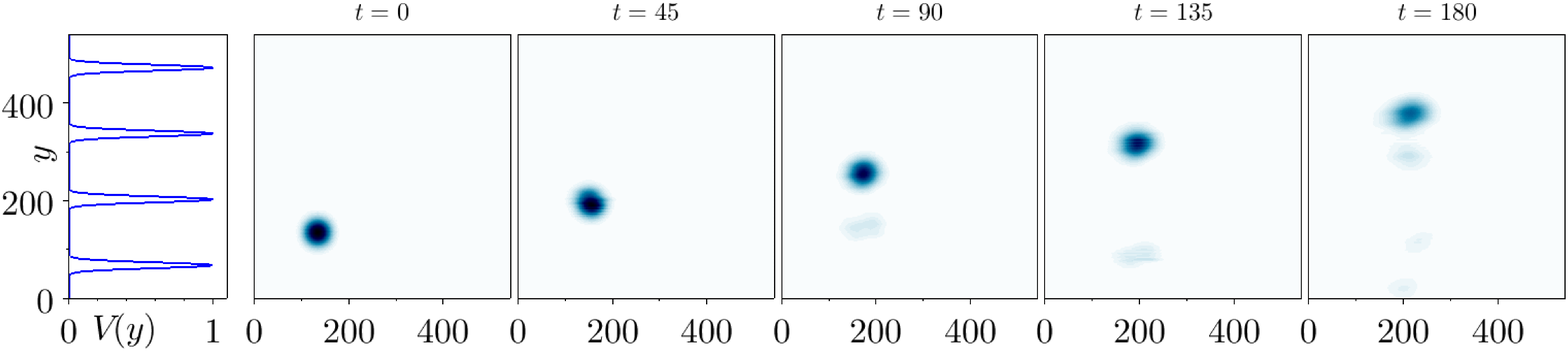}
\label{TE_barrier_Dirac_gapless}
%\ig[height=3.5cm]{edited_col_KMR_Dirac_C1_gapless.eps}}
\ig[height=3.5cm]{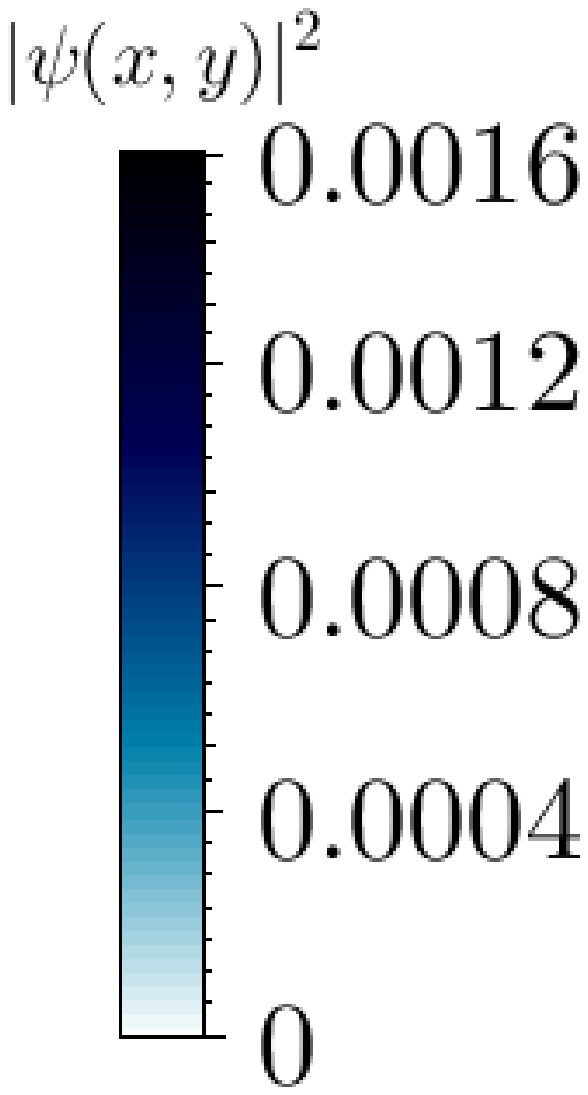}}
\end{center}
\caption{Evolution of a Gaussian wave packet in graphene with $\De_{KM} = 
\lam_R = 0.1 \ga$, and equally spaced barriers with strengths $C = \ga$ and 
spacing $d=135a$. The $x$ and $y$ coordinates (horizontal and vertical
directions respectively) are in units of $a$ while $t$ is in units of 
$\hbar /\ga$.
In momentum space the wave packet is centered around $\vk_0 = (4\pi/
(3\sqrt{3}a),\pi/(5a))$, i.e., close to a Dirac point. Since $\De_{KM} = 
\lam_R$, we have both gapless and gapped Dirac states. In (a) we see that the 
gapped mode almost completely reflects back and forth between two successive 
barriers (a small amount of transmission is faintly visible), whereas in (b) 
the gapless mode Klein tunnels through the barriers (a small amount of
reflection is visible).}
\end{figure}
\end{center}
\end{widetext}

\subsection{Effect of spin-orbit couplings}

We finally consider the case when both Kane-Mele and Rashba SO couplings are 
present and are of equal strength, i.e., $\De_{KM} = \lam_R$. As discussed in 
Sec. II and shown in Fig.~\ref{fig01}, the dispersion in this case has both 
gapped and gapless states close to the Dirac point. We look at these two kinds
of states separately. In both cases we start with a wave packet with width 
$W_x = W_y = 16a$ and peak momentum $\vk_0 = (4\pi/(3\sqrt{3}a),\pi/(5a))$. 
If the initial wave packet is constructed from the gapped states which have a 
non-relativistic dispersion, we find that there is almost complete reflection 
from the barriers. As shown in Fig.~\ref{TE_barrier_Dirac_gapped} the wave 
packet is trapped between two barriers, each of height $C=1$. The gapless 
mode however has a massless relativistic dispersion and just Klein tunnels 
through these barriers. Figure~\ref{TE_barrier_Dirac_gapless} depicts this 
case. We see that a small amount of reflection occurs when the wave packet 
crosses the barrier. This is because, as in Fig.~\ref{TE_barrier_Dirac_gapped},
we have taken the peak momentum to be at $\vk_0 = (4\pi/(3\sqrt{3}a),\pi/(5a))$
which is slightly away from the Dirac point lying at $K = (4\pi/(3\sqrt{3}a),
0)$; hence the Klein tunneling is not perfect. (Note that this wave packet 
is at normal incidence in the continuum language because the deviation of 
$\vk_0$ from $K$ is zero in the $x$-direction).

\section{Discussion}

In this paper we have studied the effects of SO couplings and
a periodic potential on the dispersion and wave packet dynamics of 
electrons in graphene. We have considered both Kane-Mele and Rashba SO 
couplings and have shown that they have interesting effects, particularly
when their magnitudes are equal.

We have first considered the continuum theory around the Dirac points
to study the effects of a periodic potential. While a periodic potential 
is known to generate new Dirac points, we have shown that SO couplings 
generally open gaps at those points. However, when the Kane-Mele and Rashba 
SO couplings are equal in magnitude, some of the gapless Dirac points are 
restored. We have shown analytically that this occurs because equal 
Kane-Mele and Rashba SO couplings produce two kinds of states, with massless 
Dirac and massive Dirac forms respectively; at normal incidence, the massless 
states transmit perfectly through an arbitrary potential, and therefore no 
gaps are generated at the ends of the Brillouin zone when a periodic 
potential is present. Next, we have used a lattice model to study the effect 
of a single potential barrier. Using the momentum along the barrier as a 
good quantum number effectively reduces the system to a one-dimensional 
lattice. We have shown that the energy spectrum obtained using the lattice
model reproduces those found with the continuum theory. In addition, we
find some additional states which are localized along the barrier. These
states have an interesting spin and sublattice structure arising from the
SO couplings. Finally, we have used the lattice model to study the time
evolution of a wave packet; the wave packet is taken to be a Gaussian.
Without the SO couplings, we discover that
there are six points in the momentum space such that a wave packet centered
around these points shows almost no spreading; we call these the no-spreading
points and we identify them by the condition that all the second derivatives
of the energy with respect to the momenta should be zero. In the absence
of SO couplings, we show that a wave packet centered around a Dirac point 
Klein tunnels through a barrier at normal incidence as expected. In the 
presence of equal Kane-Mele and Rashba SO couplings, we show that the 
massless Dirac states Klein tunnels at normal incidence while the massive
Dirac states reflect when the barrier is high.

The no-spreading points lie at an energy of $\ga \simeq 2.8$ ~eV which is 
quite far from the Dirac points (i.e., the Fermi energy of undoped graphene). 
It is therefore not easy to access them experimentally. One way of studying 
the dynamics at such points may be to inject an electron with that energy
at one point of the system and then measure the probability of detecting it 
at another point. However, such an experiment may be difficult to perform
because the large distance from the Fermi energy implies that the lifetime 
of the electron would be small. Even if it is difficult to study the 
no-spreading points in the immediate future, we have discussed them in this
paper because they are so unusual. While no-spreading points are not
uncommon in one-dimensional systems, graphene is the only example of a
two-dimensional system that we know of which has such no-spreading points.

Our results can be tested experimentally by preparing samples of graphene with
strong SO couplings. While the intrinsic SO coupling in graphene is very weak,
one can induce SO couplings in a variety of ways~\cite{weeks11,kou13,zhang14,
zollner15}, and the strength of the induced SO couplings can be tuned 
experimentally. For instance, Ref.~\onlinecite{gmitra09} shows using a first 
principles calculation that the Kane-Mele and Rashba SO couplings can be made 
equal by applying a transverse electric field equal to $2.44$ V/nm.
Finally, our work may also be applicable to other two-dimensional 
materials like silicene, germanene and stanene whose lattice structures are 
similar to graphene but with intrinsic spin-orbit couplings which are much 
stronger than in graphene~\cite{liu11,rachel14}.

\section*{Acknowledgments}

We thank Amit Agarwal, Anindya Das, Supriyo Datta and Paritosh Karnatak
for interesting discussions. D.S. thanks DST, India for Project 
No. SR/S2/JCB-44/2010 for financial support.

\end{document}